\documentclass[a4paper,twocolumn]{revtex4-1}

\usepackage{bm,amsmath}
\usepackage{graphicx}
\usepackage{color}
\usepackage{gensymb}
\usepackage[utf8]{inputenc}
\usepackage{mathrsfs}
\usepackage{bbold}
\usepackage{hyperref}

\usepackage{xcolor}
\hypersetup{
 colorlinks,
 linkcolor={red!100!black},
 citecolor={blue!100!green},
 urlcolor={blue!100!black}
 }

\newcommand{\p}{\partial}
\newcommand{\f}{\frac}
\renewcommand{\d}{\mathrm{d}}
\renewcommand{\a}{\alpha}
\newcommand{\tg}{\tilde{\gamma}}
\renewcommand{\l}{\lambda}

\newcommand{\br}{{\bf{r}}}

\begin{document}

\title{Activity-dependent self-regulation of viscous length scales in biological systems}

\author{Saroj Kumar Nandi}
\email{saroj.nandi@weizmann.ac.il}
\affiliation{Max-Planck Institute f{\"{u}}r Physik Komplexer Systeme, Dresden, Germany}
\affiliation{Department of Chemical and Biological Physics, Weizmann Institute of Science, Rehovot - 7610001, Israel}
\altaffiliation{Current address}

\begin{abstract}
Cellular cortex, which is a highly viscous thin cytoplasmic layer just below the cell membrane, controls the cell's mechanical properties, which can be characterized by a hydrodynamic length scale $\ell$. Cells actively regulate $\ell$ via the activity of force generating molecules, such as myosin II.
Here we develop a general theory for such systems through coarse-grained hydrodynamic approach including activity in the static description of the system providing an experimentally accessible parameter and elucidate the detailed mechanism of how a living system can actively self-regulate its hydrodynamic length scale, controlling the rigidity of the system. 
Remarkably, we find that $\ell$, as a function of activity, behaves universally and roughly inversely proportional to the activity of the system. Our theory rationalizes a number of experimental findings on diverse systems and comparison of our theory with existing experimental data show good agreement.
\end{abstract}
\maketitle

\section{Introduction}
The mechanical properties of a cell are governed by a highly-viscous thin layer of cytoplasm, the cell-cortex, just below the cell membrane. For most practical purposes, cortex can be assumed consisting of actin filaments and myosin motors. Cells must actively regulate the mechanical properties of the cortex that can be characterized by a hydrodynamic length scale, $\ell=\sqrt{\eta/\gamma}$, where $\eta$ is the viscosity of the cortex and $\gamma$ is some friction (e.g., exerted by the cell cytoplasm and cell membrane on the cortex). For example, $\ell$ is much bigger in the quiescent state compared to that when the cell is dividing \cite{cellbook,mayer2010,turlier2014}. Apart from its essential role in the growth and development of living organisms, understanding the detailed behavior of $\ell$ is important from other aspects such as the efficient design of artificial tissues, cellular self-assembly as well as in the field of Biomechanics \cite{huang1999,jakab2004,rumi2007}. It is known that the activity of the motor molecules, like myosin and kinesin, change the property of an assembly of filament-like molecules and motor molecules. A number of recent studies have explored the effect of activity on the properties of such systems \cite{ahmed2015,loftus2014,angelini2011,koenderink2009,zhou2009,fodor2015,gladrow2016,battle2016,sheinman2015,alvarado2013}, for example, myosin II activity softens suspended cells \cite{chan2015}, changes the viscoelastic behavior of an actomyosin network \cite{humphrey2002} and controls the dynamics in mouse oocytes \cite{almonacid2015}, bacterial cytoplasm (devoid of myosin) gets fluidized by metabolic activity \cite{parry2014}, kinesin motors lead to spontaneous flow in an assembly of microtubules \cite{sanchez2012} etc. The existing theories of actomyosin cortex have mostly focused on the quiescent state and treated the system as an elastic network \cite{koenderink2009,wang2011,wang2012,gladrow2016}. In this work, we are interested in the dynamic state when the cell is about to undergo a division \cite{turlier2014,sundar2014} and a quantitative description of how activity affects $\ell$ in this regime does not yet exist.

Developing a theory of a highly viscous system, such as cell cortex or a dense assembly of actin and myosin molecules, requires the application of glassy phenomenology \cite{parry2014,zhou2009,nishizawa2017}. Coarse-grained hydrodynamic approaches, including the complexity of the system through the effective parameters of the theory for a minimal description, have been immensely successful in describing active systems \cite{turlier2014,sebastian2012,sundar2014,sriramrmp,ranft2010,cates2015,jacques15}; we extend such theories to their dense regime. The existing coarse-grained approaches assume a spatial variation in active stress coming from the spatial variation of motor activity and the resulting force transmission \cite{mayer2010,sundar2014}. However, these models typically assume an effective viscosity reflecting the internal dissipation and successfully reproduce the the spatiotemporal cortical dynamics during cell division and polarization \cite{turlier2014,mayer2010,sundar2014}.

On the other hand, molecular approaches, taking the chemical details of rapid actin turnover through polymerization and depolymerization, their remodeling and force generation mechanism through cross-linking and myosin motors activity \cite{robin2014,fritzsche2016,carlsson2010} have predicted patterns of cortical flow. It remains unclear how to connect this microscopic origin of force generation to large-scale cortical properties. In an important recent work \cite{william2017}, McFadden {\it et al} have developed a minimal 2$D$ model of cross-linked actin filaments and myosin motors providing a framework to understand the effects of local microscopic modulation of interactions on the large-scale cellular flows.

In this work, we first develop a generic theory for an assembly of fibrillar and active force generating molecules in the dense high-viscous regime, applicable for the systems of our interest, through the approach of coarse-grained hydrodynamics. We treat the dynamics of the system within the mode-coupling theory (MCT) framework of glassy systems \cite{das2004,goetzebook}. In this work, we concentrate on the effect of myosin activity on the hydrodynamic length scale, $\ell$, and show that biological systems can actively self-regulate their length scales controlling their rigidity generalizing recent experimental findings \cite{humphrey2002,parry2014,chan2015,almonacid2015,sanchez2012}. 
We further find that $\ell$ is roughly inversely proportional to activity irrespective of the nature of the system. This is an important, testable prediction of the theory and comparison of our theory with existing experimental data shows good agreement. The rest of the paper is organized as follows: We provide a description of the system of our interest and the simplifying assumptions in Sec. \ref{description}. The detailed theory is presented in Sec. \ref{theory} followed by the results in Sec. \ref{results}. We conclude the paper by discussing our results in Sec. \ref{disc}. We have relegated some of the technical details to an Appendix.

\begin{figure}
 \includegraphics[width=\columnwidth]{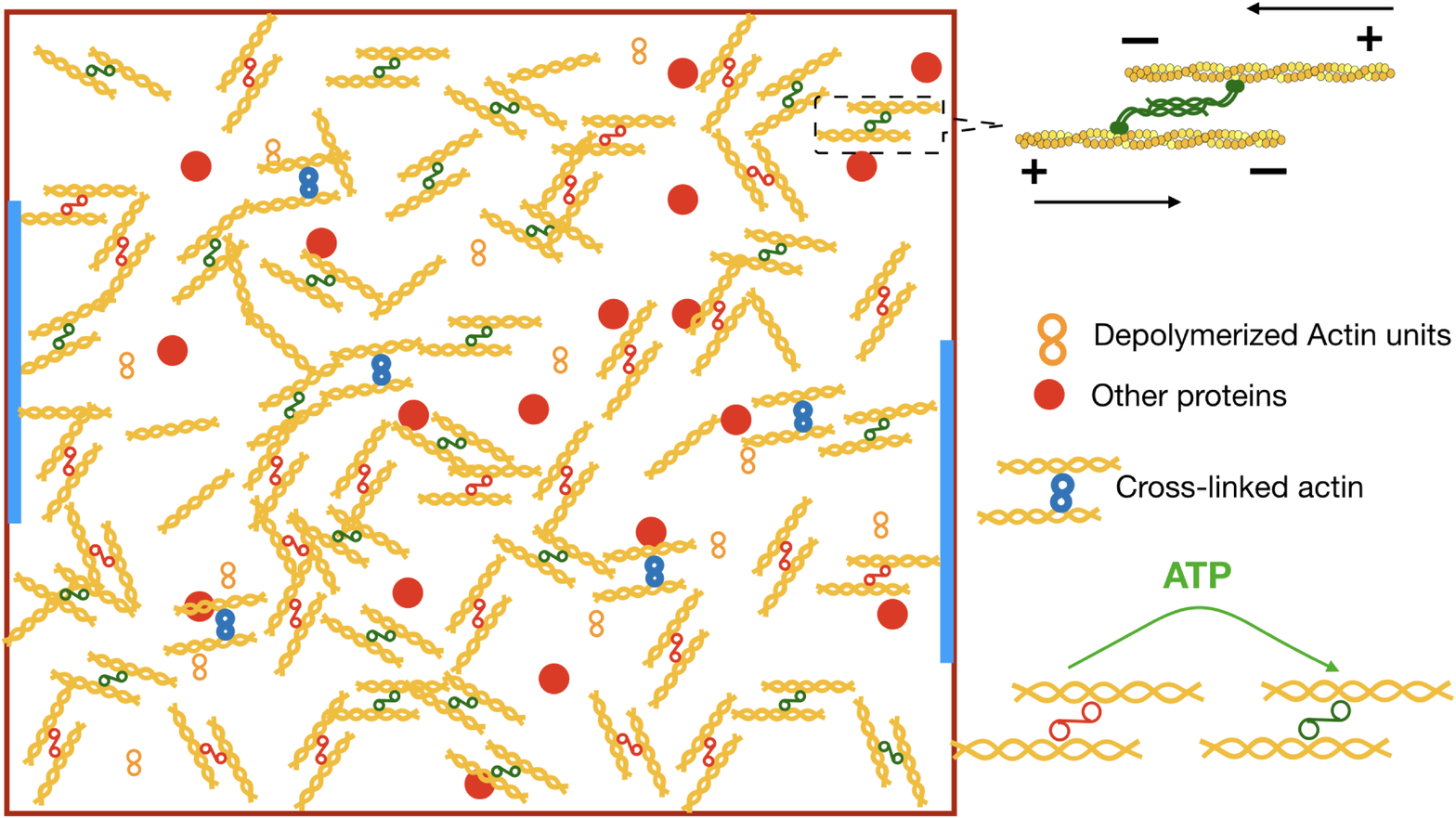}
 \caption{A simplified illustration of the system of our interest. It can be a part of a cell cortex as in \cite{chan2015,almonacid2015} or an in vitro assembly of fibrillar molecules like actin or microtubule and force generators like myosin or kinesin \cite{humphrey2002,sanchez2012}. The force generating molecules can be either in active or inactive states. When in the active state, as shown at the top in the right, they work both as cross-linkers as well as force generators by walking on two actin filaments. Myosin molecules walk from $-$ to $+$ ends of actin filaments, pulling the filaments towards each other, thus generating force in the network. On the other hand, when myosin molecules are in the inactive state, they act only as cross-linkers. Myosin molecules can change their states from the inactive to the active state through ATP consumption. There are also other cross-linking proteins, like $\alpha$-actinin, in the cortex. The blue patches are representative of focal adhesions (not to the scale, as they are 2-3 orders of magnitude larger than the dimensions of individual proteins). Actin filaments, through cross-linking and attachment with the focal adhesion points create a matrix of inhomogeneous density. All other proteins are treated as passive particles whereas myosin molecules in the active state constitute the active particles in the system.} 
 \label{schematicfluid}
\end{figure}

\section{Description of the system}
\label{description}
For concreteness, we discuss the model in terms of the actomyosin network in the cortex of cells as pictorially illustrated in Fig. \ref{schematicfluid}, however, the theory is applicable to any assembly of filament-like molecules and the active force generators in the assembly. Cortex is a complicated, dynamic assembly of many proteins; here we consider a simplified picture to enable theoretical development. We set up the problem by first discussing the individual microscopic mechanisms below.

\begin{itemize}
 \item 
 {\bf Turnover dynamics of actin filaments:} The actin filaments within cortex are dynamic in nature, constantly being polymerized at one end and depolymerized at the other \cite{fritzsche2013,fritzsche2016,cellbook}. We assume that the turnover dynamics is in a steady state and is not affected by the activity of myosin motor molecules. We thus describe the system through a time-independent inhomogeneous density modulation and a fluctuating part of density that includes the effect of turnover dynamics as well as dynamics of other molecules (Fig. \ref{schematicfluid}).

\item
 {\bf Cross-linkers:} Several types of molecules, such as $\alpha$-actinin as well as myosin molecules act as cross-linkers among actin filaments \cite{fritzsche2013,cellbook}. We do not differentiate between the cross-linking abilities of different molecules. Cross-linkers help to entangle the filament-like molecules and is parametrized by the inhomogeneous density within our theory.

 \item
{\bf Focal adhesions:} 
The actin filament network is also attached to the focal adhesion points (Fig. \ref{schematicfluid}). Since we are not interested in the effects of external medium, we simply assume that ATP concentration does not change the effects of focal adhesions, which simply contribute to the inhomogeneous density within the cortex.

\item
 {\bf Force generators in the cortex:} Myosin II molecules walk on two actin filaments and generate forces in the cortex. Since we are interested in the general picture here, and not in the detailed chemistry, from now on we denote the force generators simply as myosin. Actin filaments are polar molecules and myosin walks from the negative to the positive ends (Fig. \ref{schematicfluid}) thus pulling the two filaments towards each other. Myosin molecules can be in two different states, active and inactive; they generate forces when they are in the active state and act as cross-linkers alone when inactive. We designate the myosin molecules in the active state as the active particles. The myosin molecules can interchange their states of either active or inactive through ATP consumption.

 With a fixed amount of ATP supply, we can assume on the average a constant number of myosin molecules in the active state. Thus, the amount of ATP and myosin molecules in the active state are proportional within our theory.
\end{itemize}

To be specific, force generation in the cortex is proportional to the gradient of myosin density. However, local inhomogeneity of active myosin molecules can also lead to forces. In the experiments, one measure the amount of myosin and their specific state remains unknown \cite{mayer2010}. We assume the force generation is proportional to the amount of active myosin, and thus, proportional to ATP concentration. There can also be a rearrangement of the polarity of the actin filaments \cite{daniel2016} or disassembly of actin filaments due to some enzyme \cite{annececile2012,haviv2008} that we do not consider in this work for simplicity.

Thus, our theoretical model consists of the following: the filament-like molecules along with the cross-linkers lead to a static inhomogeneous density and there is a time-dependent density fluctuation due to turnover dynamics as well as dynamics of other proteins (that we take as passive particles) in the cortex (Fig. \ref{schematicfluid}. ATP concentration controls the fraction of myosin molecules being in the active state, which is proportional to force generation in the system. The cell boundary as well as cytoplasm impart a friction force on the cortex.

\section{Theory}
\label{theory}
Including all the considerations discussed in the previous section, we write the total density of the system as $\rho({\bf r},t)=\rho_0+\delta m({\bf r})+\delta\rho({\bf r},t)$ at position $\br$ and time $t$ where $\rho_0$ is a uniform density, $\delta m({\bf r})$, the static inhomogeneous density modulation and $\delta\rho({\bf r},t)$ is the fluctuating part of density. Note that $\rho_0$ has contributions from active and inactive particles as well as the actin filaments. $\delta m(\br)$ has contribution only due to actin filaments and this is small due to the dynamic nature of the filaments. There is some evidence that myosin II acts as actin depolymerization factor when the activity of myosin molecules exceed a certain (unknown) threshold \cite{haviv2008,annececile2012}; this may also contribute to reducing $\delta m(\br)$, that leads to fluidization of the assembly. However, we expect this effect to be small and there should always be a static inhomogeneous component of the density in the cell in the time-scale of our interest. Through a detailed microscopic calculation, Wang {\it et al} \cite{wangJCP2011} have also shown that one must keep the inhomogeneous static density component for the description of such systems.
Then, we have the continuity equation for density as
\begin{equation}
\label{continuity}
 \f{\p\rho({\bf r},t)}{\p t}=-\nabla\cdot[\rho({\bf r},t){\bf v}({\bf r},t)]
\end{equation}
where ${\bf v}({\bf r},t)$ is the velocity of a hydrodynamic volume element of the cortical constituents. The continuity equation for the momentum density, $\rho(\br,t){\bf v}(\br,t)$, is
\begin{align}
\label{ns}
 \f{\p(\rho{\bf v})}{\p t}+\nabla\cdot(\rho{\bf v}{\bf v})=&\eta\nabla^2{\bf v}+(\zeta+\eta/3)\nabla\nabla\cdot{\bf v} \nonumber\\
 &-\gamma {\bf v}-\rho\nabla\f{\delta F^U}{\delta\rho}+{\bf f},
\end{align}
where $\eta$ and $\zeta$ are shear and bulk viscosities of the fluid, ${\bf f}$, a non-conserving Gaussian white noise, with zero mean and the statistics $\langle{\bf f}_i(\br,t){\bf f}_j(\br',t')\rangle=-2k_BT[\eta \nabla^2 \mathbb{I}+(\zeta+\eta/3)\nabla \nabla]\delta(\br-\br')\delta(t-t')\delta_{ij}$ where $k_B$ is the Boltzmann constant, $T$, the temperature, $\mathbb{I}$, the identity matrix and $\delta_{ij}$, the Kronecker delta-function. 
$\gamma$ is a friction coefficient that encodes the friction force exerted by both the cytoplasm as well as the cell membrane on the cortex. $F^U$ is a suitably defined energy functional that includes the effect of matrix density of the system. Note that we use the term `matrix' to refer to the static inhomogeneous density in the system.

Active systems are inherently out of equilibrium; within MCT for active systems, it has been shown that this nonequilibrium nature of the system is manifested through an evolving effective temperature, $T_{eff}(t)$, defined through a generalized fluctuation-dissipation relation as a function of time, $t$ \cite{noneqmct}. $T_{eff}(t)$ is same as the equilibrium temperature at a very short time and dynamically evolves to a larger value, determined by the parameters of activity, at long times; note that $t$ refers to the time difference of the two-time correlation functions \cite{noneqmct}. We find that the dynamics of the system can be understood through the long-time limit of $T_{eff}(t\to\infty)\equiv T_{eff}$. A number of other studies have also shown the role of effective temperature in the dynamics of active systems \cite{shen2004,lu2006,wang2011,wang2013,loi2011,loi2011a,cugliandolo2011,suma2014,benisaac2015}. We propose to include activity in the static description of the system and this provides an experimentally accessible parameter for direct comparison with experiments on such systems.
We consider a system of $N$ active particles with the inter-particle interaction potential as $\mathcal{V}({\bf r}_1,\ldots,{\bf r}_N)$:
\begin{equation}
 \mathcal{V}=\f{1}{2}\sum_{j,l}v(|{\bf r}_j(t)-{\bf r}_l(t)|)
\end{equation}
where ${\bf r}_j(t)$ is the position of $j$th particle at time $t$. The density at ${\bf r}$ at time $t$ is defined as $\rho({\bf r},t)=\sum_{j=1}^N\delta({\bf r}-{\bf r}_j(t))$, which we can write in Fourier space as
\begin{equation}
 \rho_{\bf k}(t)=\sum_{j=1}^N e^{i{\bf k}\cdot{\bf r}_j(t)}
\end{equation}
where ${\bf k}$ is the wave vector. Following Ref. \cite{zaccarelli2002} we write
\begin{align}
 \ddot{\rho}_{\bf k}(t)=&-\sum_j({\bf k}\cdot\dot{\bf r}_j(t))^2e^{i{\bf k}\cdot{\bf r}_j(t)} \nonumber\\
 &-\f{1}{V}\sum_{{\bf k}'}v_{{\bf k}'}({\bf k}\cdot{\bf k}')\rho_{{\bf k}-{\bf k}'}(t)\rho_{{\bf k}'}(t)
\end{align}
where we have set the mass of the particles to unity and $v_{\bf k}$ is the Fourier components of the interaction potential. Following Zwanzig \cite{zwanzigbook}, we write this equation as
\begin{equation}\label{zwanzigform}
 \ddot{\rho}_{\bf k}(t)+\Omega_{\bf k}\rho_{\bf k}(t)=\mathcal{F}_{\bf k}(t).
\end{equation}
The frequency term $\Omega_{\bf k}$ that minimizes the residual forces $\mathcal{F}_{\bf k}(t)$ is given as
\begin{equation}
\label{omega}
 \Omega_{\bf k}=\f{k^2}{S_k^A}k_B(T+T_{e})
\end{equation}
where we have used the relation $\langle \dot{r}_{i\alpha}\dot{r}_{j\beta}\rangle \sim k_B(T+T_e)\delta_{ij}\delta_{\alpha\beta}$ for an active fluid at an effective temperature $T_{eff}=T+T_e$, where $T_e$ is the excess contribution coming from activity alone and $S_k^A$ is the equal-time two-point correlation function for such a system
\begin{equation}
 S_k^A=\f{1}{N}\langle \rho_k(t)\rho_{-k}(t)\rangle.
\end{equation}
Then we obtain the residual forces from Eq. (\ref{zwanzigform}) as
\begin{align}
\label{residual_f}
 \mathcal{F}_{\bf k}(t)=\Omega_k \rho_k(t)-\sum_j({\bf k}\cdot\dot{\bf r}_j(t))^2e^{i{\bf k}\cdot{\bf r}_j(t)} \nonumber\\
 -\f{1}{V}\sum_{{\bf k}'}v_{{\bf k}'}({\bf k}\cdot{\bf k}')\rho_{{\bf k}-{\bf k}'}(t)\rho_{{\bf k}'}(t).
\end{align}
The fact that the residual forces are minimized with respect to $\Omega_k$ implies an orthogonality condition
\begin{equation}
 \langle \rho_{-k}(t)\mathcal{F}_k(t)\rangle=0.
\end{equation}
Using the detailed form of residual forces from Eq. (\ref{residual_f}), we obtain the equation for $c_k^A$, which is analogous to the direct correlation function of a passive system,
\begin{align}
\label{active_c}
 c_k^A=-\f{\beta^A}{k^2N^2S_k^A}\sum_{{\bf k}'}v_{{\bf k}'}({\bf k}\cdot{\bf k}')\langle\rho_{-{\bf k}}(t)\rho_{{\bf k}-{\bf k}'}(t)\rho_{{\bf k}'}(t)\rangle,
\end{align}
where $c_k^A$ is defined through the relation $c_k^A=(1-1/S_k^A)/\rho_0$ with $\rho_0$ being the average density of the system and
\begin{equation}
 \beta^A=\f{1}{k_B(T+T_e)}\simeq \f{1}{k_BT}(1-T_e/T)\simeq \f{1}{k_BT}f(A),
\end{equation}
where we have written the contribution from activity in $f(A)$ denoting activity as $A$. Note that $\beta^A$, in principle, could depend on wave vector, however, we expect $T_e$ at all wave vectors to be the same as the behavior of the correlation and response functions at all wave vectors within MCT is similar \cite{noneqmct} though we do not have an {\it a priori} justification for this. If the activity is zero, we have $f(A)=1$ and Eq. (\ref{active_c}) gives $c_k$, the direct correlation function of the corresponding passive system. Thus, we can write
\begin{equation}\label{active_c2}
 c_k^A=f(A)c_k.
\end{equation}
An intuitive way to understand Eq. (\ref{active_c2}) is as follows: $c_k$ is related to the inter-atomic interaction potential. Consider one of the particles in the system, this particle interacts with an effective potential obtained from integrating out the other particles. Activity introduces a random motion of the particle, considering the activity of different particles uncorrelated. Since the state of myosin locally depends on the availability of ATP, such an assumption is justified. Thus, activity reduces correlation in the particle motion and $f(A)$ is a monotonically decreasing function of $A$. When the activity is small, we can write $f(A)=e^{-A}$.
We note here that it is not clear what should be the actual form of $c^A(r)$. Even in an equilibrium fluid, it is not clear what should enter as $c(r)$ in the Ramakrishnan-Yussouff free energy functional \cite{ramakrishnan1979} in the dynamical mean-field theory for density waves. Moreover, Eq. (\ref{active_c}), as well as its equilibrium counterpart, is only valid within the random phase approximation \cite{zaccarelli2002,saroj2015}. However, this difficulty is not important in the context of the present work as we do not use $c_k^A$ directly within the theory. As we are interested in an assembly of active and passive particles embedded in a static inhomogeneous density background (that we call the matrix), $A$ is directly proportional to the amount of active particles or the amount of ATP \cite{benisaac2015}. Similar use of effective parameters to describe an active network material, within the random first-order transition theory framework has been demonstrated elsewhere \cite{wangJCP2011,wang2012,wang2013}.

The velocity field in a biological system is small and we ignore the second term in Eq. (\ref{ns}). 
We linearize Eqs. (\ref{continuity}) and (\ref{ns}), take the divergence of the second and eliminate velocity using the first to obtain
\begin{align}
 \f{\p \delta\rho({\bf r},t)}{\p t}&=D_L\nabla^2\f{\p\delta\rho({\bf r},t)}{\p t}-\f{\gamma}{\rho_0}\f{\p \delta\rho({\bf r},t)}{\p t} \nonumber\\
 &+\nabla\cdot\left[\rho\nabla\f{\delta F^U[\delta\rho({\bf r},t),\delta m]}{\delta\rho({\bf r},t)}\right]-\nabla\cdot{\bf f}
\end{align}
where $D_L=(\zeta+4\eta/3)/\rho_0$. The energy functional $F^U$ for the active system is
\begin{align}
\beta F^U&=\int_{\bf r}\left\{\rho({\bf r},t)\ln\f{\rho({\bf r},t)}{\rho_0}-[\rho({\bf r},t)-\rho_0]\right\}  \nonumber\\
& -\f{1}{2}\int_{{\bf r},{\bf r}'}c^A(|{\bf r}-{\bf r}'|)[\rho({\bf r},t)-\rho_0][\rho({\bf r}',t)-\rho_0]  \nonumber \\
&+\beta\int_{\bf r}U({\bf r})\rho({\bf r},t)
\end{align}
where $\int_{\bf r}\equiv \int {\d {\bf r}}$. The first part is an ideal gas contribution and the second part contains the contribution due to interaction. It is not clear what should be the detailed form of $c^A(\br)$, however, this is not important for the purpose of the present work \footnote{Note that even in a passive system it is not clear what should be the correct form of $c(\br)$ (see Appendix).}. The contribution due to the inhomogeneous density modulation, $\delta m(\br)$, is encoded through a potential (see Appendix for details) $U({\bf r})$, which makes it straightforward to extend the theory for systems under external force, for example, a cell attached to an elastomeric membrane that can be stretched as in Ref. \cite{jungbauer2008}.

To characterize the dynamics of a dense system, we need to look at the two-point correlation functions \cite{hansenmcdonald}. This is a directly accessible quantity in experiments and all the transport coefficients can be calculated from this function through the application of statistical mechanics. We define the two-point correlation function $S_{k}(t)=\langle\delta\rho_{k}(0)\delta\rho_{-{k}}(t)\rangle$ and the normalized two-point correlation function  $\phi_{k}(t)=S_{k}(t)/S_{k}(t=0)$. Then the equation for $\phi_k(t)$ becomes (see Appendix for details)
\begin{align}
 \f{\p^2\phi_k(t)}{\p t^2}&+\Gamma_k\f{\p\phi_k(t)}{\p t}+\f{k_BTk^2}{S_k^{A}}\phi_k(t) \nonumber\\
 &+\int_0^t\mathcal{M}_k(t-t')\f{\p\phi_k(t')}{\p t'}\d t'=0
\end{align}
where $\Gamma_k=D_Lk^2+\gamma/\rho_0$ and $\mathcal{M}_k(t)$ is the memory kernel given as
\begin{align} \label{memoryk}
 \mathcal{M}_k(t)&=\f{k_BT\rho_0}{2k^2}\int_{\bf q}[{\bf k}\cdot{\bf q}c_q^A+{\bf k}\cdot({\bf k}-{\bf q})c_{k-q}^A]^2S_{k-q}(t)S_q(t) \nonumber\\
 &+\f{k_BT\rho_0}{k^2}\int_{\bf q}\left[{\bf k}\cdot{\bf q}c_q^A+\f{{\bf k}\cdot({\bf k}-{\bf q})}{\rho_0}\right]^2S_{k-q}^bS_q(t)
\end{align}
where $S_k^b=\langle\delta m_k\delta m_{-k}\rangle$. Activity in our model enters through $c_k^A$ within the memory kernel and $S_k^A$ which is defined as $1/(1-\rho_0 c_k^A)$.

The complexity of the full wavevector dependent theory hinders its applicability to a biological system. To render simpler form we take a schematic approximation suppressing the wavevector dependence leaving a single-mode, however, retaining the essential mechanism of the theory \cite{leutheusser1984,kirkpatrick1985,brader2009}. Within this approximation, we write the theory for one wave vector $k_{max}$ that corresponds the first maximum of structure factor \footnote{Within MCT, the relaxation dynamics at all wave vectors are similar \cite{das2004}, therefore, writing the theory for any other wave vector is equivalent.}. Ignoring the acceleration term, which is not important at long times and for systems of our interest, we obtain the equation of motion for the normalized two-time correlation function, $\phi(t)\equiv \phi_{k=k_{max}}(t)$, in this limit as
\begin{equation}\label{schematicmct}
 (1+\tg)\f{\p\phi(t)}{\p t}+\Omega\phi(t)+\int_0^t\mathcal{M}(t-t')\f{\p\phi(t')}{\p t'}\d t'=0
\end{equation}
where $\tilde{\gamma}=\gamma/\rho_0D_Lk_{max}^2$ and we have ignored the activity dependence in the frequency term $\Omega$. The memory kernel becomes
\begin{equation}\label{schematicmemory}
 \mathcal{M}(t)\simeq\left(\lambda_1^0-\a\sqrt{\l_1^0}\right)\phi(t)+\left(\lambda_2^0-\a\sqrt{\l_2^0}\right)\phi(t)^2.
\end{equation}
The main contribution in the dynamics comes from the parts inside the integrals in Eq. (\ref{memoryk}). Therefore, $\l_1^0$ is proportional to the square of the matrix density (through $S_{k_{max}}^b$) and both $\l_1^0$ and $\l_2^0$ are proportional to average density $\rho_0$ and inversely proportional to temperature (see Appendix, Eq. (A8), for their detailed forms).
The parameter $\a$ contains the information of activity of the system and $\a \propto A $ is proportional to the amount of active myosin or ATP concentration. $\l_1^0$ contains the information of matrix density as well as the interaction of the fluid with the matrix and $\l_2^0$ encodes the strength of interaction within the fluid. Both $\l_1^0$ and $\l_2^0$ increases as the density of the fluid increases or temperature decreases and only $\l_1^0$ increases as the matrix density increases.

\begin{figure}
\includegraphics[width=\columnwidth]{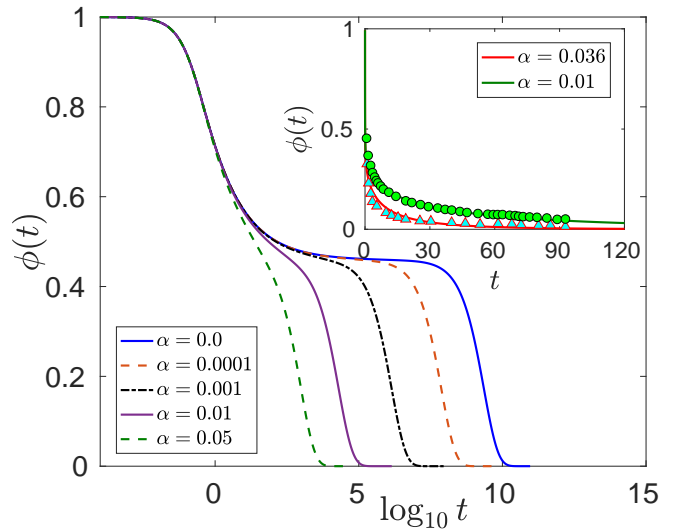}
\caption{We show the decay of the correlation function $\phi(t)$ as a function of time obtained through the numerical solution of Eqs. (\ref{schematicmct}) and (\ref{schematicmemory}). $\phi(t)$ shows a fast decay to a plateau at short times and then a much slower decay to zero at longer times. Activity, $\a$, makes the decay of $\phi(t)$ faster. We have used the parameters as $\tilde{\gamma}=1$, $\l_1^0=0.28$ and $\l_2^0=3.417$. {\bf Inset:} We compare our theory with the experimental data of Ref. \cite{almonacid2015} for the dynamics of cytoplasm in mouse oocytes. The data for $\phi(t)$ is taken from Fig. 4(b) of Ref. \cite{almonacid2015} and solid lines are fit of our theory showing that activity makes the dynamics faster as $\phi(t)$ decays quicker at larger $\a$. The parameters $\l_1^0=0.8$ and $\l_2^0=1.98$ are same for both the curves.}
\label{active_corrfn}
\end{figure}

\section{Results}
\label{results}
The theory, Eqs. (\ref{schematicmct}) and (\ref{schematicmemory}) for a passive system when $\a=0$ is well-understood in the literature \cite{sing1998,goetze2000,goetzebook,das2004}. The correlation function $\phi(t)$ decays to zero quite fast at small density when $\l_1^0$ and $\l_2^0$ are small and develops a two-step relaxation scenario at higher density; it first decays to a plateau and then to zero from the plateau at long times. At even larger density, $\phi(t)$ does not decay to zero anymore, showing an ergodicity breaking transition:
a discontinuous transition at $\l_2^0=4$ when $\l_1^0=0$ and a continuous transition at $\l_1^0=1$ when $\l_2^0=0$ \cite{goetzebook,das2004}. It is well-known that this transition shifts to higher density (larger $\l_1^0$ or $\l_2^0$) in active systems of self-propelled particles \cite{berthier2013,mandal2016,flenner2016,noneqmct,viscmct}, our theory also has similar feature as can be easily seen from Eq. (\ref{schematicmemory}).
The ergodicity breaking transition is a failure of the theory whose origin is not yet well-understood \cite{rfimmct} and all the predictions of the theory break down beyond this point.
In the context of the systems of our interest, the transition by itself is not important as living systems can not, in general, go to a state with very high density or low temperature. We expect our theory to work within the regime where biological systems operate.

We set $\tilde{\gamma}$ to unity and show the decay of $\phi(t)$ for $\l_1^0=0.28$ and $\l_2^0=3.417$ for different values of the activity parameter $\a$ in Fig. \ref{active_corrfn}. The correlation function decays faster as the activity becomes stronger, that is at higher values of $\a$. Ref. \cite{almonacid2015} looked at the cytoplasmic dynamics for positioning the nucleus in mouse oocytes. We have obtained the data for correlation function, as presented in Fig. 4(b) in \cite{almonacid2015} and fit our theory with the data with $\l_1^0=0.8$ and $\l_2^0=1.98$ (same for both curves) and $\a=0.01$ and $0.036$ for the two sets of data as shown in the inset of Fig. \ref{active_corrfn}. Note that there is a difference in the scale of measuring times in the theory and the experiment; we have rescaled the time in our calculation by a factor of $0.01$ to match with the experimental data. The important point in this fit is that higher activity fluidizes the cytoplasm and the correlation function decays faster.

\begin{figure}
\includegraphics[width=8.6cm]{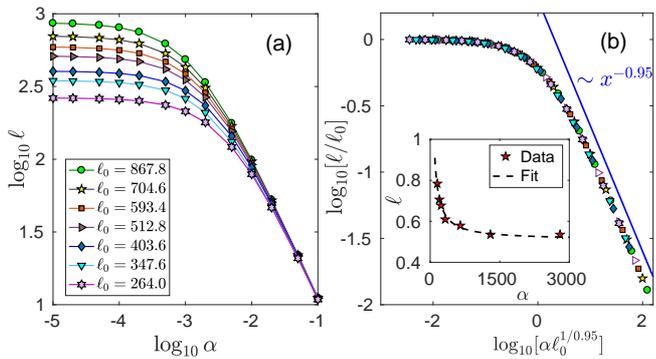}
\caption{ (a) Different systems are characterized by different values of $(\l_1^0,\l_2^0)$ leading to distinct values of $\ell_0$. We show the behavior of $\ell$ for many such systems as a function of $\a$. 
(b) Since there is a characteristic length scale, $\ell_0$, independent of activity, we expect a scaling for $\ell$ as $\ell=\ell_0^\theta f(\alpha\ell_0^{1/\nu})$ and obtain data collapse for $\theta=1.0$ and $\nu=0.95$ implying $\ell\sim 1/\a^{0.95}$ in a regime where activity dominates the dynamics. {\bf Inset:} We obtain the data of length scale from Ref. \cite{sanchez2012} (see text). Our theory shows $\ell\sim 1/\a$; therefore we fit the data with $f(x)=a+b/x$. We find excellent agreement between theory and experiment with $a=0.51$ and $b=40.13$.}
\label{visclength}
 \end{figure}

We now show how activity controls the hydrodynamic length scale $\ell$ of the system, constituting the main result of this work. We use Kubo formula to obtain the viscosity from the correlation function. The renormalized viscosity of the system is given by $ \eta_R=\int_0^\infty\phi(t')\d t'$ \cite{kubo1966,hansenmcdonald}. This relation is valid for systems in equilibrium. However, considering that the system is in a state where linear-response theory is valid, we can still use Kubo formula for the viscosity.
The hydrodynamic length scale, which determines the rigidity of the system is defined, following Refs. \cite{mayer2010,sundar2014}, as $\ell=\sqrt{\eta_R/\tilde{\gamma}}\sim \sqrt{\eta_R}$ since we have set $\tilde{\gamma}$ to unity. 
For a passive system, viscosity, and hence the hydrodynamic length scale, is determined by $(\l_1^0,\l_2^0)$ alone. However, in an active system viscosity depends on activity; as activity sets in, the renormalized viscosity is different from its passive value.

Different biological systems are characterized by distinct values of $(\l_1^0,\l_2^0)$ as the number of proteins and molecules attached to the focal adhesion points are different and the densities of the fluid components of the systems also vary. Correspondingly, they can be characterized by distinct $\ell_0 =\sqrt{\eta_R(\alpha=0)}$. 
We choose a particular set of values for $(\l_1^0,\l_2^0)$ and denote the system with the corresponding passive length scale $\ell_0$. We show the behavior of $\ell$ as a function of activity for different $\ell_0$ in Fig. \ref{visclength}(a). $\ell$ saturates at small $\a$ to a value corresponding to its passive value $\ell_0$ and decreases as $\a$ increases.
Thus, biological systems can actively self-regulate their hydrodynamic length scales, hence, rigidity by changing their activity.

Since there is a characteristic length scale $\ell_0$, independent of activity for each system, we expect a scaling relation \cite{amitbook} for $\ell$ as $\ell=\ell_0^\theta f(\alpha\ell_0^{1/\nu})$ and obtain scaling collapse for $\theta=1$ and $\nu=0.95$ as shown in Fig. \ref{visclength}(b). From this we obtain 
\begin{equation}
 \ell\sim 1/\a
\end{equation}
in the regime where activity dominates the dynamics. This implies that irrespective of the nature of the system, the viscous length scale $\ell$ follows a simple algebraic law as a function of the activity parameter $\alpha$. As we define $\ell\sim\sqrt{\eta_R}$, the viscosity behaves as $\sim 1/\a^{1.9}$ as a function of activity in the activity dominated regime.

For a comparison with experiment, we look at the data of velocity correlation function, $\langle V(R)V(0)\rangle$, as a function of distance $R$ as presented in the inset of Fig. 2(c) in \cite{sanchez2012}. At a certain $R$, $\langle V(R)V(0)\rangle$ changes because of activity. The correlation in a system is controlled by certain length scales. We can assume this change in correlation at a fixed $R=R_0$ varies exponentially: $[\langle V(R)V(0)\rangle]_{R_0} \sim e^{1/\ell}$; this gives $\ell$ up to a constant factor. We have extracted $\ell$ in this way from \cite{sanchez2012} for $R_0=12$ (the first data point in each curve). The data are presented in terms of ATP concentration; within our theory, this is proportional to the active molecules, designated by the parameter $\a$. A fit of this data with our theory, presented in the inset of Fig. \ref{visclength}(b) shows excellent agreement. Ref. \cite{humphrey2002} has shown that myosin molecules in the active state fluidizes an in vitro actomyosin gel and reduces the local shear relaxation time, \cite{parry2014} has shown that bacterial cytoplasm becomes solid-like in the dormant state and fluidizes when metabolically active and \cite{chan2015} has shown that myosin activity softens cells in suspension. Our theory rationalizes these findings in terms of the activity of constituent molecules.

\section{Discussion}
\label{disc}
Our theory is within the framework of mode-coupling theory (MCT) for dense super-cooled systems \cite{das2004,goetzebook}. MCT provides a mechanism of the rapid increase of viscosity and relaxation time of systems as temperature decreases or density increases. We have extended this theory to include activity generated by some active motor proteins like myosin or kinesin. Considering detailed microscopic dynamics, Ref. \cite{gladrow2016} shows that mode-coupling naturally emerges in the dynamics of active networks. MCT shows an ergodicity-breaking transition above certain values of the parameters and the theory fails to describe the system beyond this point \cite{giulioreview,goetzebook,das2004}. This transition is not important for biological systems as living systems cannot become extremely dense; we expect the theory to be valid in the regime where most biological systems operate. Our theory should be viewed as a description, starting from the molecular level, of the mechanism how motor activity changes the dynamics at the macroscopic length scale. Our theory is similar in spirit to the microscopic theory developed by Wolynes {\it et al} for such systems \cite{wangJCP2011,wang2012,wang2013}.

We have implemented activity in the static description of the system, where activity parameter $\alpha$ is proportional to the ATP concentration. Thus, our theory allows a direct experimentally accessible parameter.
The principal result of this work is the behavior of the viscous length scale $\ell$, which is an important quantity as it controls the mechanical properties of a system, as a function of activity; we find that $\ell$ is roughly inversely proportional to the amount of active myosin, generalizing recent experimental findings \cite{humphrey2002,parry2014,almonacid2015,chan2015,sanchez2012} on such systems. Ref. \cite{zhou2009} shows that viscosity of a cell increases with ATP depletion, our theory is in qualitative agreement with this finding. Remarkably, we find universality in the behavior of $\ell$, which varies roughly inversely with $a$ irrespective of the nature of the system. This is an important result readily testable in experiments and simulations. Comparison of our theory with the existing experimental results show good agreement (Fig. 3b). The increased viscosity as well as $\ell^2$ in ATP-depleted nucleoli in {\it Xenopus laevis} oocytes \cite{brangwynne2011} are also in qualitative agreement with our predictions.

The cellular cortex is a complex system, where a large number of active processes control the large-scale behavior. We have considered a simplified picture ignoring a number of such processes. For example, we have neglected the chiral nature of actin-filaments \cite{sundar2014,sebastian2013} in our theory, however, we believe, including chirality within the theory is not going to affect the main conclusions of this work. 
We have treated the effects of cross-linking from non-force-generating molecules, like the $\alpha$-actinin, and that of myosin molecules to be same. It is important to understand how ATP concentration affects the cross-linking that we have modeled through the static inhomogeneous density $\delta m$. Detailed molecular modeling along the lines of Ref. \cite{william2017} should unravel this aspect.

There may as well have a number of other effects like the organization of the polarity of actin filaments \cite{daniel2016} or change in the actin polymerization dynamics due to some enzyme that are also going to affect $\ell$. We have not included such dynamics in this work, however, we expect these effects to give higher order corrections in the behavior of $\ell$. The effect of myosin activity where myosin acts as actin depolymerization factor \cite{haviv2008} when the activity of myosin crosses a certain threshold value is going to fluidize the system. Although we did not look at this effect in detail, this is already included in the coarse-grained description of reducing correlation within our theory. 
As the length scale directly gives the rigidity of the system, a study along the lines of \cite{chan2015} or \cite{humphrey2002} as a function of active myosin molecules of the system provides a straightforward way to further test our theory.

We thank Ludovic Berthier, Anagha Datar, Stephan Grill, Frank J{\"{u}}licher, Sriram Ramaswamy, Nir S. Gov, Grzegorz Szamel, Daniel Riveline and Jacques Prost for many important and enlightening discussions and Koshland Foundation for funding through a fellowship.

\appendix

\section{Some details of the MCT calculation}
Here we first show how the energy functional $F^U$, and hence the force density, can be written in terms of the static inhomogeneous density $\delta m({\bf r})$.
The density fluctuates around the static density $m({\bf r})=\rho_0+\delta m({\bf r})$. Minimizing $F^U$ with respect to $m({\bf r})$ and replacing $U({\bf r})$ in terms of $m({\bf r})$, we obtain the force density as
\begin{align}
 \rho\nabla\f{\delta \beta F^U}{\delta\rho({\bf r},t)}=&\nabla\int_{{\bf r}'}[\delta({\bf r}-{\bf r}')-\rho_0c^A(|{\bf r}-{\bf r}'|)]\delta\rho({\bf r}',t)\nonumber\\
-&\nabla\int_{{\bf r}'}\delta m({\bf r})c^A(|{\bf r}-{\bf r}'|)\delta\rho({\bf r}',t) \nonumber\\
-&\f{\nabla m({\bf r})}{m({\bf r})}\int_{{\bf r}'}[\delta({\bf r}-{\bf r}')-m({\bf r})c^A(|{\bf r}-{\bf r}'|)]\delta\rho({\bf r}',t)\nonumber\\
-&\delta\rho({\bf r},t)\nabla\int_{{\bf r}'}c^A(|{\bf r}-{\bf r}'|)\delta\rho({\bf r}',t).
\end{align}

Defining Fourier transform of different quantities as $\delta\rho({\bf r},t)=\int_{\bf k}\delta\rho_k(t)e^{i{\bf k}\cdot{\bf r}}$, where we have used the notation $\int_{\bf k}\equiv\int \f{\d^dk}{(2\pi)^d}$ in $d$-dimension, we can write the equation of motion for the density fluctuation in Fourier space as
 \begin{align}
 \label{eqofm1}
  \left[ \f{\p^2}{\p t^2}+\Gamma_k\f{\p}{\p t}+\f{k_BTk^2}{S_k^{A}}\right]&\delta\rho_k(t)=-i{\bf k}\cdot[{\bf F}_{\bf k}+{\bf f}_{\bf k}] \nonumber\\
  & \hspace{-1cm}=-i{\bf k}\cdot[{\bf F}_{\bf k}^{m\rho}+{\bf F}_{\bf k}^{\rho\rho}+{\bf f_k}],
 \end{align}
where $\Gamma_k=D_Lk^2+\gamma/\rho_0$ and $S_k^{A}=1/(1-\rho_0c^A_k)$. The Force densities are given as
\begin{align}
\label{sm:eqofm2}
 {\bf F}_{\bf k}^{m\rho}(t)&=ik_BT\int_{\bf q}[{\bf k}c^A_{\bf q}+({\bf k}-{\bf q})/\rho_0S_{\bf q}^{A}]\delta m_{{\bf k}-{\bf q}}\delta\rho_{\bf q}(t) \\
 {\bf F}_{\bf k}^{\rho\rho}(t)&=\f{i}{2}k_BT\int_{\bf q}[{\bf q}c^A_{\bf q}+({\bf k}-{\bf q})c^A_{{\bf k}-{\bf q}}]\delta\rho_{\bf q}(t)\delta\rho_{{\bf k}-{\bf q}}(t) \label{sm:eqofm3}.
\end{align}
In this work, we have included activity in $c_k^A$ and $S_k^A$ and ${\bf f}_{\bf k}$ is the bare noise, in contrast to the existing MCT theories on active systems of self-propelled particles \cite{szamel2016,noneqmct}. Using the standard prescription of MCT through hydrodynamic approach \cite{kawasaki2002,saroj2011} to extract the excess damping due to interactions in terms of a memory kernel given as
\begin{equation}
\mathcal{M}_{\bf k}(t)=\f{1}{k_BTV}\langle {\bf F}_{\bf k}(0)\cdot{\bf F}_{-{\bf k}}(t)\rangle,
\end{equation}
where $V$ is the system volume and ${\bf F}_{\bf k}(t)$ is given by Eqs. (\ref{sm:eqofm2}) and (\ref{sm:eqofm3}). 
The two-point correlation function is defined as $S_{k}(t)=\langle\delta\rho_{k}(0)\delta\rho_{-{k}}(t)\rangle$. The normalized two-point correlation function is $\phi_{k}(t)=S_{k}(t)/S_{k}(t=0)$ and we obtain the mode-coupling equation as
\begin{align}\label{sm:mcteq}
 \f{\p^2\phi_k(t)}{\p t^2}&+\Gamma_k\f{\p\phi_k(t)}{\p t}+\f{k_BTk^2}{S_k^{A}}\phi_k(t) \nonumber\\
 &+\int_0^t \mathcal{M}_k(t-t')\f{\p\phi_k(t')}{\p t'}\d t'=0
\end{align}
where $\mathcal{M}_k(t)$ is the memory kernel given as
\begin{align} \label{sm:memoryk_SM}
 \mathcal{M}_k(t)&=\f{k_BT\rho_0}{2k^2}\int_{\bf q}[{\bf k}\cdot{\bf q}c_q^A+{\bf k}\cdot({\bf k}-{\bf q})c_{k-q}^A]^2S_{k-q}(t)S_q(t) \nonumber\\
 &+\f{k_BT\rho_0}{k^2}\int_{\bf q}\left[{\bf k}\cdot{\bf q}c_q^A+\f{{\bf k}\cdot({\bf k}-{\bf q})}{\rho_0}\right]^2S_{k-q}^bS_q(t)
\end{align}
with $S_k^b=\langle\delta m_k\delta m_{-k}\rangle$. The theory in this form is too complex to be applicable to the systems of our interest as there are too many unknown parameters. Therefore, we take a schematic approximation, which amounts to writing the theory for only one wave vector, $k_{max}$. This approximation provides a simpler form for better understanding and applicability, keeping the key mechanism of the theory intact \cite{das2004,goetzebook,brader2009}. We divide Eq. (\ref{sm:mcteq}) by $D_L k^2$ and write $\tilde{\gamma}=\gamma/(\rho_0D_L k^2)$. We expand $c_k^A\sim c_k(1-A)$ (Eq. \ref{active_c2}) and ignore the terms second order in $A$ while completing the squares in Eq. (\ref{sm:memoryk_SM}). The first part of $\mathcal{M}_k(t)$ contains two factors of $\phi(t)\equiv \phi_{k_{max}}(t)$ and the second part contains one $\phi(t)$. The time-independent parts in Eq. (\ref{sm:memoryk_SM}) are defined as $\l_2^0$ and $\l_1^0$ respectively:
\begin{align}
 \l_2^0 &\equiv\left[\f{k_BT\rho_0}{2D_Lk^4}\int_{\bf q}[{\bf k}\cdot{\bf q}c_q^A+{\bf k}\cdot({\bf k}-{\bf q})c_{k-q}^A]^2S_{k-q}S_q\right]_{k_{max}}\nonumber\\
 \l_1^0 &\equiv \left[\f{k_BT\rho_0}{D_Lk^4}\int_{\bf q}\left[{\bf k}\cdot{\bf q}c_q^A+\f{{\bf k}\cdot({\bf k}-{\bf q})}{\rho_0}\right]^2S_{k-q}^bS_q\right]_{k_{max}},
\end{align}
where $S_k\equiv S_k(t=0)$ and $[\ldots]_{k_{max}}$ implies that we need to evaluate this term at this wave vector. Then the linear order terms in $A$ in the first and second parts of the memory kernel are proportional to $\a\sqrt{\l_2^0}$ and $\a\sqrt{\l_1^0}$ respectively where $\a$ is proportional to $A$. Note that the main contributions in the dynamics come from the terms under the integrals in Eq. (\ref{sm:memoryk_SM}) and not from the prefactors. Thus, we obtain the memory kernel under the schematic approximation as
\begin{equation}\label{sm:schematicmemory}
\mathcal{M}(t)\simeq(\l_1^0-\a\sqrt{\l_1^0})\phi(t)+(\l_2^0-\a\sqrt{\l_2^0})\phi^2(t)
\end{equation}
with the mode coupling equation for $\phi(t)$ obtained as
\begin{equation}\label{sm:schematicmct}
  (1+\tg)\f{\p\phi(t)}{\p t}+\Omega\phi(t)+\int_0^t\mathcal{M}(t-t')\f{\p\phi(t')}{\p t'}\d t'=0
\end{equation}
where we have neglected the acceleration term in Eq. (\ref{sm:mcteq}). $\Omega\equiv k_BT/D_L S_k^A$ and we neglect activity dependence in $\Omega$ as it doesn't affect the dynamics in a strong way. We need to solve Eq. (\ref{sm:schematicmct}) along with Eq. (\ref{sm:schematicmemory}) numerically where we have set $\Omega$ and $\tilde{\gamma}$ to unity redefining the time scales. This is the reason why the time-scales in our theory and that in Ref. \cite{almonacid2015} are different and we need to rescale the time-scale in our calculation for comparison with the experimental data.


\begin{thebibliography}{76}%
\makeatletter
\providecommand \@ifxundefined [1]{%
 \@ifx{#1\undefined}
}%
\providecommand \@ifnum [1]{%
 \ifnum #1\expandafter \@firstoftwo
 \else \expandafter \@secondoftwo
 \fi
}%
\providecommand \@ifx [1]{%
 \ifx #1\expandafter \@firstoftwo
 \else \expandafter \@secondoftwo
 \fi
}%
\providecommand \natexlab [1]{#1}%
\providecommand \enquote  [1]{``#1''}%
\providecommand \bibnamefont  [1]{#1}%
\providecommand \bibfnamefont [1]{#1}%
\providecommand \citenamefont [1]{#1}%
\providecommand \href@noop [0]{\@secondoftwo}%
\providecommand \href [0]{\begingroup \@sanitize@url \@href}%
\providecommand \@href[1]{\@@startlink{#1}\@@href}%
\providecommand \@@href[1]{\endgroup#1\@@endlink}%
\providecommand \@sanitize@url [0]{\catcode `\\12\catcode `\$12\catcode
  `\&12\catcode `\#12\catcode `\^12\catcode `\_12\catcode `\%12\relax}%
\providecommand \@@startlink[1]{}%
\providecommand \@@endlink[0]{}%
\providecommand \url  [0]{\begingroup\@sanitize@url \@url }%
\providecommand \@url [1]{\endgroup\@href {#1}{\urlprefix }}%
\providecommand \urlprefix  [0]{URL }%
\providecommand \Eprint [0]{\href }%
\providecommand \doibase [0]{http://dx.doi.org/}%
\providecommand \selectlanguage [0]{\@gobble}%
\providecommand \bibinfo  [0]{\@secondoftwo}%
\providecommand \bibfield  [0]{\@secondoftwo}%
\providecommand \translation [1]{[#1]}%
\providecommand \BibitemOpen [0]{}%
\providecommand \bibitemStop [0]{}%
\providecommand \bibitemNoStop [0]{.\EOS\space}%
\providecommand \EOS [0]{\spacefactor3000\relax}%
\providecommand \BibitemShut  [1]{\csname bibitem#1\endcsname}%
\let\auto@bib@innerbib\@empty
\bibitem [{\citenamefont {Alberts}\ \emph {et~al.}(2002)\citenamefont
  {Alberts}, \citenamefont {Johnson}, \citenamefont {Lewis}, \citenamefont
  {Raff}, \citenamefont {Roberts},\ and\ \citenamefont {Walter}}]{cellbook}%
  \BibitemOpen
  \bibfield  {author} {\bibinfo {author} {\bibfnamefont {B.}~\bibnamefont
  {Alberts}}, \bibinfo {author} {\bibfnamefont {A.}~\bibnamefont {Johnson}},
  \bibinfo {author} {\bibfnamefont {J.}~\bibnamefont {Lewis}}, \bibinfo
  {author} {\bibfnamefont {M.}~\bibnamefont {Raff}}, \bibinfo {author}
  {\bibfnamefont {K.}~\bibnamefont {Roberts}}, \ and\ \bibinfo {author}
  {\bibfnamefont {P.}~\bibnamefont {Walter}},\ }\href@noop {} {\emph {\bibinfo
  {title} {Molecular Biology of the Cell}}},\ \bibinfo {edition} {4th}\ ed.\
  (\bibinfo  {publisher} {Garland Science, New York},\ \bibinfo {year}
  {2002})\BibitemShut {NoStop}%
\bibitem [{\citenamefont {Mayer}\ \emph {et~al.}(2010)\citenamefont {Mayer},
  \citenamefont {Depken}, \citenamefont {Bois}, \citenamefont
  {J{\"{u}}licher},\ and\ \citenamefont {Grill}}]{mayer2010}%
  \BibitemOpen
  \bibfield  {author} {\bibinfo {author} {\bibfnamefont {M.}~\bibnamefont
  {Mayer}}, \bibinfo {author} {\bibfnamefont {M.}~\bibnamefont {Depken}},
  \bibinfo {author} {\bibfnamefont {J.~S.}\ \bibnamefont {Bois}}, \bibinfo
  {author} {\bibfnamefont {F.}~\bibnamefont {J{\"{u}}licher}}, \ and\ \bibinfo
  {author} {\bibfnamefont {S.~W.}\ \bibnamefont {Grill}},\ }\href {\doibase
  10.1038/nature09376} {\bibfield  {journal} {\bibinfo  {journal} {Nature}\
  }\textbf {\bibinfo {volume} {467}},\ \bibinfo {pages} {617} (\bibinfo {year}
  {2010})}\BibitemShut {NoStop}%
\bibitem [{\citenamefont {Turlier}\ \emph {et~al.}(2014)\citenamefont
  {Turlier}, \citenamefont {Audoly}, \citenamefont {Prost},\ and\ \citenamefont
  {Joanny}}]{turlier2014}%
  \BibitemOpen
  \bibfield  {author} {\bibinfo {author} {\bibfnamefont {H.}~\bibnamefont
  {Turlier}}, \bibinfo {author} {\bibfnamefont {B.}~\bibnamefont {Audoly}},
  \bibinfo {author} {\bibfnamefont {J.}~\bibnamefont {Prost}}, \ and\ \bibinfo
  {author} {\bibfnamefont {J.-F.}\ \bibnamefont {Joanny}},\ }\href {\doibase
  10.1016/j.bpj.2013.11.014} {\bibfield  {journal} {\bibinfo  {journal}
  {Biophys. J.}\ }\textbf {\bibinfo {volume} {106}},\ \bibinfo {pages} {114}
  (\bibinfo {year} {2014})}\BibitemShut {NoStop}%
\bibitem [{\citenamefont {Huang}\ and\ \citenamefont
  {Ingber}(1999)}]{huang1999}%
  \BibitemOpen
  \bibfield  {author} {\bibinfo {author} {\bibfnamefont {S.}~\bibnamefont
  {Huang}}\ and\ \bibinfo {author} {\bibfnamefont {D.~E.}\ \bibnamefont
  {Ingber}},\ }\href {\doibase 10.1038/13043} {\bibfield  {journal} {\bibinfo
  {journal} {Nature Cell Biol.}\ }\textbf {\bibinfo {volume} {1}},\ \bibinfo
  {pages} {E131} (\bibinfo {year} {1999})}\BibitemShut {NoStop}%
\bibitem [{\citenamefont {Jakab}\ \emph {et~al.}(2004)\citenamefont {Jakab},
  \citenamefont {Neagu}, \citenamefont {Mironov}, \citenamefont {Markwald},\
  and\ \citenamefont {Forgacs}}]{jakab2004}%
  \BibitemOpen
  \bibfield  {author} {\bibinfo {author} {\bibfnamefont {K.}~\bibnamefont
  {Jakab}}, \bibinfo {author} {\bibfnamefont {A.}~\bibnamefont {Neagu}},
  \bibinfo {author} {\bibfnamefont {V.}~\bibnamefont {Mironov}}, \bibinfo
  {author} {\bibfnamefont {R.~R.}\ \bibnamefont {Markwald}}, \ and\ \bibinfo
  {author} {\bibfnamefont {G.}~\bibnamefont {Forgacs}},\ }\href {\doibase
  10.1073/pnas.0400164101} {\bibfield  {journal} {\bibinfo  {journal} {Proc.
  Natl. Acad. Sci.}\ }\textbf {\bibinfo {volume} {101}},\ \bibinfo {pages}
  {2864} (\bibinfo {year} {2004})}\BibitemShut {NoStop}%
\bibitem [{\citenamefont {De}\ \emph {et~al.}(2007)\citenamefont {De},
  \citenamefont {Zemel},\ and\ \citenamefont {Safran}}]{rumi2007}%
  \BibitemOpen
  \bibfield  {author} {\bibinfo {author} {\bibfnamefont {R.}~\bibnamefont
  {De}}, \bibinfo {author} {\bibfnamefont {A.}~\bibnamefont {Zemel}}, \ and\
  \bibinfo {author} {\bibfnamefont {S.~A.}\ \bibnamefont {Safran}},\ }\href
  {\doibase 10.1038/nphys680} {\bibfield  {journal} {\bibinfo  {journal} {Nat.
  Phys.}\ }\textbf {\bibinfo {volume} {3}},\ \bibinfo {pages} {655} (\bibinfo
  {year} {2007})}\BibitemShut {NoStop}%
\bibitem [{\citenamefont {Ahmed}\ \emph {et~al.}(2015)\citenamefont {Ahmed},
  \citenamefont {Fodor},\ and\ \citenamefont {Betz}}]{ahmed2015}%
  \BibitemOpen
  \bibfield  {author} {\bibinfo {author} {\bibfnamefont {W.~W.}\ \bibnamefont
  {Ahmed}}, \bibinfo {author} {\bibfnamefont {{\'{E}}.}~\bibnamefont {Fodor}},
  \ and\ \bibinfo {author} {\bibfnamefont {T.}~\bibnamefont {Betz}},\ }\href
  {\doibase 10.1016/j.bbamcr.2015.05.022} {\bibfield  {journal} {\bibinfo
  {journal} {Biochimica et Biophysica Acta}\ }\textbf {\bibinfo {volume}
  {1853}},\ \bibinfo {pages} {3083} (\bibinfo {year} {2015})}\BibitemShut
  {NoStop}%
\bibitem [{\citenamefont {Loftus}\ \emph {et~al.}(2014)\citenamefont {Loftus},
  \citenamefont {Shmygol},\ and\ \citenamefont {Richardson}}]{loftus2014}%
  \BibitemOpen
  \bibfield  {author} {\bibinfo {author} {\bibfnamefont {F.~C.}\ \bibnamefont
  {Loftus}}, \bibinfo {author} {\bibfnamefont {A.}~\bibnamefont {Shmygol}}, \
  and\ \bibinfo {author} {\bibfnamefont {M.~J.~E.}\ \bibnamefont
  {Richardson}},\ }\href {\doibase 10.1113/jphysiol.2014.275412} {\bibfield
  {journal} {\bibinfo  {journal} {J. Physil.}\ }\textbf {\bibinfo {volume}
  {592}},\ \bibinfo {pages} {4447} (\bibinfo {year} {2014})}\BibitemShut
  {NoStop}%
\bibitem [{\citenamefont {Angelini}\ \emph {et~al.}(2011)\citenamefont
  {Angelini}, \citenamefont {Hannezo}, \citenamefont {Trepat}, \citenamefont
  {Marquez}, \citenamefont {Fredberg},\ and\ \citenamefont
  {Weitz}}]{angelini2011}%
  \BibitemOpen
  \bibfield  {author} {\bibinfo {author} {\bibfnamefont {T.~E.}\ \bibnamefont
  {Angelini}}, \bibinfo {author} {\bibfnamefont {E.}~\bibnamefont {Hannezo}},
  \bibinfo {author} {\bibfnamefont {X.}~\bibnamefont {Trepat}}, \bibinfo
  {author} {\bibfnamefont {M.}~\bibnamefont {Marquez}}, \bibinfo {author}
  {\bibfnamefont {J.~J.}\ \bibnamefont {Fredberg}}, \ and\ \bibinfo {author}
  {\bibfnamefont {D.~A.}\ \bibnamefont {Weitz}},\ }\href {\doibase
  10.1016/0022-3093(91)90266-9} {\bibfield  {journal} {\bibinfo  {journal}
  {Proc. Natl. Acad. Sci. (USA)}\ }\textbf {\bibinfo {volume} {108}},\ \bibinfo
  {pages} {4717} (\bibinfo {year} {2011})}\BibitemShut {NoStop}%
\bibitem [{\citenamefont {Koenderink}\ \emph {et~al.}(2009)\citenamefont
  {Koenderink}, \citenamefont {Dogic}, \citenamefont {Nakamura}, \citenamefont
  {Bendix}, \citenamefont {MacKintosh}, \citenamefont {Hartwige}, \citenamefont
  {Stossel},\ and\ \citenamefont {Weitz}}]{koenderink2009}%
  \BibitemOpen
  \bibfield  {author} {\bibinfo {author} {\bibfnamefont {G.~H.}\ \bibnamefont
  {Koenderink}}, \bibinfo {author} {\bibfnamefont {Z.}~\bibnamefont {Dogic}},
  \bibinfo {author} {\bibfnamefont {F.}~\bibnamefont {Nakamura}}, \bibinfo
  {author} {\bibfnamefont {P.~M.}\ \bibnamefont {Bendix}}, \bibinfo {author}
  {\bibfnamefont {F.~C.}\ \bibnamefont {MacKintosh}}, \bibinfo {author}
  {\bibfnamefont {J.~H.}\ \bibnamefont {Hartwige}}, \bibinfo {author}
  {\bibfnamefont {T.~P.}\ \bibnamefont {Stossel}}, \ and\ \bibinfo {author}
  {\bibfnamefont {D.~A.}\ \bibnamefont {Weitz}},\ }\href {\doibase
  10.1073/pnas.0903974106} {\bibfield  {journal} {\bibinfo  {journal} {Proc.
  Natl. Acad. Sci. (USA)}\ }\textbf {\bibinfo {volume} {106}},\ \bibinfo
  {pages} {15192} (\bibinfo {year} {2009})}\BibitemShut {NoStop}%
\bibitem [{\citenamefont {Zhou}\ \emph {et~al.}(2009)\citenamefont {Zhou},
  \citenamefont {Trepat}, \citenamefont {Park}, \citenamefont {Lenormand},
  \citenamefont {Oliver}, \citenamefont {Mijailovich}, \citenamefont {Hardin},
  \citenamefont {Weitz}, \citenamefont {Butler},\ and\ \citenamefont
  {Fredberg}}]{zhou2009}%
  \BibitemOpen
  \bibfield  {author} {\bibinfo {author} {\bibfnamefont {E.~H.}\ \bibnamefont
  {Zhou}}, \bibinfo {author} {\bibfnamefont {X.}~\bibnamefont {Trepat}},
  \bibinfo {author} {\bibfnamefont {C.~Y.}\ \bibnamefont {Park}}, \bibinfo
  {author} {\bibfnamefont {G.}~\bibnamefont {Lenormand}}, \bibinfo {author}
  {\bibfnamefont {M.~N.}\ \bibnamefont {Oliver}}, \bibinfo {author}
  {\bibfnamefont {S.~M.}\ \bibnamefont {Mijailovich}}, \bibinfo {author}
  {\bibfnamefont {C.}~\bibnamefont {Hardin}}, \bibinfo {author} {\bibfnamefont
  {D.~A.}\ \bibnamefont {Weitz}}, \bibinfo {author} {\bibfnamefont {J.~P.}\
  \bibnamefont {Butler}}, \ and\ \bibinfo {author} {\bibfnamefont {J.~J.}\
  \bibnamefont {Fredberg}},\ }\href {\doibase 10.1073/pnas.0901462106}
  {\bibfield  {journal} {\bibinfo  {journal} {Proc. Natl. Acad. Sci. (USA)}\
  }\textbf {\bibinfo {volume} {106}},\ \bibinfo {pages} {10632} (\bibinfo
  {year} {2009})}\BibitemShut {NoStop}%
\bibitem [{\citenamefont {Fodor}\ \emph {et~al.}(2015)\citenamefont {Fodor},
  \citenamefont {Guo}, \citenamefont {Gov}, \citenamefont {Visco},
  \citenamefont {Weitz},\ and\ \citenamefont {van Wijland}}]{fodor2015}%
  \BibitemOpen
  \bibfield  {author} {\bibinfo {author} {\bibfnamefont {{\'{E}}.}~\bibnamefont
  {Fodor}}, \bibinfo {author} {\bibfnamefont {M.}~\bibnamefont {Guo}}, \bibinfo
  {author} {\bibfnamefont {N.~S.}\ \bibnamefont {Gov}}, \bibinfo {author}
  {\bibfnamefont {P.}~\bibnamefont {Visco}}, \bibinfo {author} {\bibfnamefont
  {D.~A.}\ \bibnamefont {Weitz}}, \ and\ \bibinfo {author} {\bibfnamefont
  {F.}~\bibnamefont {van Wijland}},\ }\href {\doibase
  10.1209/0295-5075/110/48005} {\bibfield  {journal} {\bibinfo  {journal}
  {Europhys. Lett.}\ }\textbf {\bibinfo {volume} {110}},\ \bibinfo {pages}
  {48005} (\bibinfo {year} {2015})}\BibitemShut {NoStop}%
\bibitem [{\citenamefont {Gladrow}\ \emph {et~al.}(2016)\citenamefont
  {Gladrow}, \citenamefont {Fakhri}, \citenamefont {MacKintosh}, \citenamefont
  {Schmidt},\ and\ \citenamefont {Broedersz}}]{gladrow2016}%
  \BibitemOpen
  \bibfield  {author} {\bibinfo {author} {\bibfnamefont {J.}~\bibnamefont
  {Gladrow}}, \bibinfo {author} {\bibfnamefont {N.}~\bibnamefont {Fakhri}},
  \bibinfo {author} {\bibfnamefont {F.}~\bibnamefont {MacKintosh}}, \bibinfo
  {author} {\bibfnamefont {C.}~\bibnamefont {Schmidt}}, \ and\ \bibinfo
  {author} {\bibfnamefont {C.}~\bibnamefont {Broedersz}},\ }\href {\doibase
  10.1103/PhysRevLett.116.248301} {\bibfield  {journal} {\bibinfo  {journal}
  {Phys. Rev. Lett.}\ }\textbf {\bibinfo {volume} {116}},\ \bibinfo {pages}
  {248301} (\bibinfo {year} {2016})}\BibitemShut {NoStop}%
\bibitem [{\citenamefont {Battle}\ \emph {et~al.}(2016)\citenamefont {Battle},
  \citenamefont {Broedersz}, \citenamefont {Fakhri}, \citenamefont {Geyer},
  \citenamefont {Howard}, \citenamefont {Schmidt},\ and\ \citenamefont
  {MacKintosh}}]{battle2016}%
  \BibitemOpen
  \bibfield  {author} {\bibinfo {author} {\bibfnamefont {C.}~\bibnamefont
  {Battle}}, \bibinfo {author} {\bibfnamefont {C.~P.}\ \bibnamefont
  {Broedersz}}, \bibinfo {author} {\bibfnamefont {N.}~\bibnamefont {Fakhri}},
  \bibinfo {author} {\bibfnamefont {V.~F.}\ \bibnamefont {Geyer}}, \bibinfo
  {author} {\bibfnamefont {J.}~\bibnamefont {Howard}}, \bibinfo {author}
  {\bibfnamefont {C.~F.}\ \bibnamefont {Schmidt}}, \ and\ \bibinfo {author}
  {\bibfnamefont {F.~C.}\ \bibnamefont {MacKintosh}},\ }\href {\doibase
  10.1126/science.aac8167} {\bibfield  {journal} {\bibinfo  {journal}
  {Science}\ }\textbf {\bibinfo {volume} {352}},\ \bibinfo {pages} {604}
  (\bibinfo {year} {2016})}\BibitemShut {NoStop}%
\bibitem [{\citenamefont {Sheinman}\ \emph {et~al.}(2015)\citenamefont
  {Sheinman}, \citenamefont {Sharma}, \citenamefont {Alvarado}, \citenamefont
  {Koenderink},\ and\ \citenamefont {MacKintosh}}]{sheinman2015}%
  \BibitemOpen
  \bibfield  {author} {\bibinfo {author} {\bibfnamefont {M.}~\bibnamefont
  {Sheinman}}, \bibinfo {author} {\bibfnamefont {A.}~\bibnamefont {Sharma}},
  \bibinfo {author} {\bibfnamefont {J.}~\bibnamefont {Alvarado}}, \bibinfo
  {author} {\bibfnamefont {G.~H.}\ \bibnamefont {Koenderink}}, \ and\ \bibinfo
  {author} {\bibfnamefont {F.~C.}\ \bibnamefont {MacKintosh}},\ }\href
  {\doibase 10.1103/PhysRevLett.114.098104} {\bibfield  {journal} {\bibinfo
  {journal} {Phys. Rev. Lett.}\ }\textbf {\bibinfo {volume} {114}},\ \bibinfo
  {pages} {098104} (\bibinfo {year} {2015})}\BibitemShut {NoStop}%
\bibitem [{\citenamefont {Alvarado}\ \emph {et~al.}(2013)\citenamefont
  {Alvarado}, \citenamefont {Sheinman}, \citenamefont {Sharma}, \citenamefont
  {MacKintosh},\ and\ \citenamefont {Koenderink}}]{alvarado2013}%
  \BibitemOpen
  \bibfield  {author} {\bibinfo {author} {\bibfnamefont {J.}~\bibnamefont
  {Alvarado}}, \bibinfo {author} {\bibfnamefont {M.}~\bibnamefont {Sheinman}},
  \bibinfo {author} {\bibfnamefont {A.}~\bibnamefont {Sharma}}, \bibinfo
  {author} {\bibfnamefont {F.~C.}\ \bibnamefont {MacKintosh}}, \ and\ \bibinfo
  {author} {\bibfnamefont {G.~H.}\ \bibnamefont {Koenderink}},\ }\href
  {\doibase 10.1038/nphys2715} {\bibfield  {journal} {\bibinfo  {journal} {Nat.
  Phys.}\ }\textbf {\bibinfo {volume} {9}},\ \bibinfo {pages} {591} (\bibinfo
  {year} {2013})}\BibitemShut {NoStop}%
\bibitem [{\citenamefont {Chan}\ \emph {et~al.}(2015)\citenamefont {Chan},
  \citenamefont {Ekpenyong}, \citenamefont {Golfier}, \citenamefont {Li},
  \citenamefont {Chalut}, \citenamefont {Otto}, \citenamefont {Elgeti},
  \citenamefont {Guck},\ and\ \citenamefont {Lautenschl{\"{a}}ger}}]{chan2015}%
  \BibitemOpen
  \bibfield  {author} {\bibinfo {author} {\bibfnamefont {C.~J.}\ \bibnamefont
  {Chan}}, \bibinfo {author} {\bibfnamefont {A.~E.}\ \bibnamefont {Ekpenyong}},
  \bibinfo {author} {\bibfnamefont {S.}~\bibnamefont {Golfier}}, \bibinfo
  {author} {\bibfnamefont {W.}~\bibnamefont {Li}}, \bibinfo {author}
  {\bibfnamefont {K.~J.}\ \bibnamefont {Chalut}}, \bibinfo {author}
  {\bibfnamefont {O.}~\bibnamefont {Otto}}, \bibinfo {author} {\bibfnamefont
  {J.}~\bibnamefont {Elgeti}}, \bibinfo {author} {\bibfnamefont
  {J.}~\bibnamefont {Guck}}, \ and\ \bibinfo {author} {\bibfnamefont
  {F.}~\bibnamefont {Lautenschl{\"{a}}ger}},\ }\href {\doibase
  10.1016/j.bpj.2015.03.009} {\bibfield  {journal} {\bibinfo  {journal}
  {Biophys. J.}\ }\textbf {\bibinfo {volume} {108}},\ \bibinfo {pages} {1856}
  (\bibinfo {year} {2015})}\BibitemShut {NoStop}%
\bibitem [{\citenamefont {Humphrey}\ \emph {et~al.}(2002)\citenamefont
  {Humphrey}, \citenamefont {Duggan}, \citenamefont {Saha}, \citenamefont
  {Smith},\ and\ \citenamefont {K{\"{a}}s}}]{humphrey2002}%
  \BibitemOpen
  \bibfield  {author} {\bibinfo {author} {\bibfnamefont {D.}~\bibnamefont
  {Humphrey}}, \bibinfo {author} {\bibfnamefont {C.}~\bibnamefont {Duggan}},
  \bibinfo {author} {\bibfnamefont {D.}~\bibnamefont {Saha}}, \bibinfo {author}
  {\bibfnamefont {D.}~\bibnamefont {Smith}}, \ and\ \bibinfo {author}
  {\bibfnamefont {J.}~\bibnamefont {K{\"{a}}s}},\ }\href {\doibase
  10.1038/416413a} {\bibfield  {journal} {\bibinfo  {journal} {Nature}\
  }\textbf {\bibinfo {volume} {416}},\ \bibinfo {pages} {413} (\bibinfo {year}
  {2002})}\BibitemShut {NoStop}%
\bibitem [{\citenamefont {Almonacid}\ \emph {et~al.}(2015)\citenamefont
  {Almonacid}, \citenamefont {Ahmed}, \citenamefont {Bussonnier}, \citenamefont
  {Mailly}, \citenamefont {Betz}, \citenamefont {Voituriez}, \citenamefont
  {Gov},\ and\ \citenamefont {Verlhac}}]{almonacid2015}%
  \BibitemOpen
  \bibfield  {author} {\bibinfo {author} {\bibfnamefont {M.}~\bibnamefont
  {Almonacid}}, \bibinfo {author} {\bibfnamefont {W.~W.}\ \bibnamefont
  {Ahmed}}, \bibinfo {author} {\bibfnamefont {M.}~\bibnamefont {Bussonnier}},
  \bibinfo {author} {\bibfnamefont {P.}~\bibnamefont {Mailly}}, \bibinfo
  {author} {\bibfnamefont {T.}~\bibnamefont {Betz}}, \bibinfo {author}
  {\bibfnamefont {R.}~\bibnamefont {Voituriez}}, \bibinfo {author}
  {\bibfnamefont {N.~S.}\ \bibnamefont {Gov}}, \ and\ \bibinfo {author}
  {\bibfnamefont {M.-H.}\ \bibnamefont {Verlhac}},\ }\href {\doibase
  10.1038/ncb3131} {\bibfield  {journal} {\bibinfo  {journal} {Nat. Cell
  Biol.}\ }\textbf {\bibinfo {volume} {17}},\ \bibinfo {pages} {470} (\bibinfo
  {year} {2015})}\BibitemShut {NoStop}%
\bibitem [{\citenamefont {Parry}\ \emph {et~al.}(2014)\citenamefont {Parry},
  \citenamefont {Surovtsev}, \citenamefont {Cabeen}, \citenamefont {O'Hern},
  \citenamefont {Dufresne},\ and\ \citenamefont {Jacobs-Wagner}}]{parry2014}%
  \BibitemOpen
  \bibfield  {author} {\bibinfo {author} {\bibfnamefont {B.~R.}\ \bibnamefont
  {Parry}}, \bibinfo {author} {\bibfnamefont {I.~V.}\ \bibnamefont
  {Surovtsev}}, \bibinfo {author} {\bibfnamefont {M.~T.}\ \bibnamefont
  {Cabeen}}, \bibinfo {author} {\bibfnamefont {C.~S.}\ \bibnamefont {O'Hern}},
  \bibinfo {author} {\bibfnamefont {E.~R.}\ \bibnamefont {Dufresne}}, \ and\
  \bibinfo {author} {\bibfnamefont {C.}~\bibnamefont {Jacobs-Wagner}},\ }\href
  {\doibase 10.1016/j.cell.2013.11.028} {\bibfield  {journal} {\bibinfo
  {journal} {Cell}\ }\textbf {\bibinfo {volume} {156}},\ \bibinfo {pages} {183}
  (\bibinfo {year} {2014})}\BibitemShut {NoStop}%
\bibitem [{\citenamefont {Sanchez}\ \emph {et~al.}(2012)\citenamefont
  {Sanchez}, \citenamefont {Chen}, \citenamefont {DeCamp}, \citenamefont
  {Heymann},\ and\ \citenamefont {Dogic}}]{sanchez2012}%
  \BibitemOpen
  \bibfield  {author} {\bibinfo {author} {\bibfnamefont {T.}~\bibnamefont
  {Sanchez}}, \bibinfo {author} {\bibfnamefont {D.~T.~N.}\ \bibnamefont
  {Chen}}, \bibinfo {author} {\bibfnamefont {S.~J.}\ \bibnamefont {DeCamp}},
  \bibinfo {author} {\bibfnamefont {M.}~\bibnamefont {Heymann}}, \ and\
  \bibinfo {author} {\bibfnamefont {Z.}~\bibnamefont {Dogic}},\ }\href
  {\doibase 10.1038/nature11591} {\bibfield  {journal} {\bibinfo  {journal}
  {Nature}\ }\textbf {\bibinfo {volume} {491}},\ \bibinfo {pages} {431}
  (\bibinfo {year} {2012})}\BibitemShut {NoStop}%
\bibitem [{\citenamefont {Wang}\ and\ \citenamefont
  {Wolynes}(2011)}]{wang2011}%
  \BibitemOpen
  \bibfield  {author} {\bibinfo {author} {\bibfnamefont {S.}~\bibnamefont
  {Wang}}\ and\ \bibinfo {author} {\bibfnamefont {P.~G.}\ \bibnamefont
  {Wolynes}},\ }\href {\doibase 10.1063/1.3624753} {\bibfield  {journal}
  {\bibinfo  {journal} {J. Chem. Phys.}\ }\textbf {\bibinfo {volume} {135}},\
  \bibinfo {pages} {051101} (\bibinfo {year} {2011})}\BibitemShut {NoStop}%
\bibitem [{\citenamefont {Wang}\ and\ \citenamefont
  {Wolynes}(2012)}]{wang2012}%
  \BibitemOpen
  \bibfield  {author} {\bibinfo {author} {\bibfnamefont {S.}~\bibnamefont
  {Wang}}\ and\ \bibinfo {author} {\bibfnamefont {P.~G.}\ \bibnamefont
  {Wolynes}},\ }\href {\doibase 10.1063/1.3702583} {\bibfield  {journal}
  {\bibinfo  {journal} {J. Chem. Phys.}\ }\textbf {\bibinfo {volume} {136}},\
  \bibinfo {pages} {145102} (\bibinfo {year} {2012})}\BibitemShut {NoStop}%
\bibitem [{\citenamefont {Naganathan}\ \emph {et~al.}(2014)\citenamefont
  {Naganathan}, \citenamefont {F{\"{u}}rthauer}, \citenamefont {Nishikawa},
  \citenamefont {J{\"{u}}licher},\ and\ \citenamefont {Grill}}]{sundar2014}%
  \BibitemOpen
  \bibfield  {author} {\bibinfo {author} {\bibfnamefont {S.~R.}\ \bibnamefont
  {Naganathan}}, \bibinfo {author} {\bibfnamefont {S.}~\bibnamefont
  {F{\"{u}}rthauer}}, \bibinfo {author} {\bibfnamefont {M.}~\bibnamefont
  {Nishikawa}}, \bibinfo {author} {\bibfnamefont {F.}~\bibnamefont
  {J{\"{u}}licher}}, \ and\ \bibinfo {author} {\bibfnamefont {S.~W.}\
  \bibnamefont {Grill}},\ }\href {\doibase 10.7554/eLife.04165} {\bibfield
  {journal} {\bibinfo  {journal} {eLife}\ }\textbf {\bibinfo {volume} {3}},\
  \bibinfo {pages} {e04165} (\bibinfo {year} {2014})}\BibitemShut {NoStop}%
\bibitem [{\citenamefont {Nishizawa}\ \emph {et~al.}(2017)\citenamefont
  {Nishizawa}, \citenamefont {Fujiwara}, \citenamefont {Ikenaga}, \citenamefont
  {Nakajo}, \citenamefont {Yanagisawa},\ and\ \citenamefont
  {Mizuno}}]{nishizawa2017}%
  \BibitemOpen
  \bibfield  {author} {\bibinfo {author} {\bibfnamefont {K.}~\bibnamefont
  {Nishizawa}}, \bibinfo {author} {\bibfnamefont {K.}~\bibnamefont {Fujiwara}},
  \bibinfo {author} {\bibfnamefont {M.}~\bibnamefont {Ikenaga}}, \bibinfo
  {author} {\bibfnamefont {N.}~\bibnamefont {Nakajo}}, \bibinfo {author}
  {\bibfnamefont {M.}~\bibnamefont {Yanagisawa}}, \ and\ \bibinfo {author}
  {\bibfnamefont {D.}~\bibnamefont {Mizuno}},\ }\href {\doibase
  10.1038/s41598-017-14883-y} {\bibfield  {journal} {\bibinfo  {journal} {Sci.
  Rep.}\ }\textbf {\bibinfo {volume} {7}},\ \bibinfo {pages} {15143} (\bibinfo
  {year} {2017})}\BibitemShut {NoStop}%
\bibitem [{\citenamefont {F{\"{u}}rthauer}\ \emph {et~al.}(2012)\citenamefont
  {F{\"{u}}rthauer}, \citenamefont {Neef}, \citenamefont {Grill}, \citenamefont
  {Kruse},\ and\ \citenamefont {J{\"{u}}licher}}]{sebastian2012}%
  \BibitemOpen
  \bibfield  {author} {\bibinfo {author} {\bibfnamefont {S.}~\bibnamefont
  {F{\"{u}}rthauer}}, \bibinfo {author} {\bibfnamefont {M.}~\bibnamefont
  {Neef}}, \bibinfo {author} {\bibfnamefont {S.~W.}\ \bibnamefont {Grill}},
  \bibinfo {author} {\bibfnamefont {K.}~\bibnamefont {Kruse}}, \ and\ \bibinfo
  {author} {\bibfnamefont {F.}~\bibnamefont {J{\"{u}}licher}},\ }\href
  {\doibase 10.1088/1367-2630/14/2/023001} {\bibfield  {journal} {\bibinfo
  {journal} {New J. Phys}\ }\textbf {\bibinfo {volume} {14}},\ \bibinfo {pages}
  {023001} (\bibinfo {year} {2012})}\BibitemShut {NoStop}%
\bibitem [{\citenamefont {Marchetti}\ \emph {et~al.}(2013)\citenamefont
  {Marchetti}, \citenamefont {Joanny}, \citenamefont {Ramaswamy}, \citenamefont
  {Liverpool}, \citenamefont {Prost}, \citenamefont {Rao},\ and\ \citenamefont
  {Simha}}]{sriramrmp}%
  \BibitemOpen
  \bibfield  {author} {\bibinfo {author} {\bibfnamefont {M.~C.}\ \bibnamefont
  {Marchetti}}, \bibinfo {author} {\bibfnamefont {J.~F.}\ \bibnamefont
  {Joanny}}, \bibinfo {author} {\bibfnamefont {S.}~\bibnamefont {Ramaswamy}},
  \bibinfo {author} {\bibfnamefont {T.~B.}\ \bibnamefont {Liverpool}}, \bibinfo
  {author} {\bibfnamefont {J.}~\bibnamefont {Prost}}, \bibinfo {author}
  {\bibfnamefont {M.}~\bibnamefont {Rao}}, \ and\ \bibinfo {author}
  {\bibfnamefont {R.~A.}\ \bibnamefont {Simha}},\ }\href {\doibase
  10.1103/RevModPhys.85.1143} {\bibfield  {journal} {\bibinfo  {journal} {Rev.
  Mod. Phys.}\ }\textbf {\bibinfo {volume} {85}},\ \bibinfo {pages} {1143}
  (\bibinfo {year} {2013})}\BibitemShut {NoStop}%
\bibitem [{\citenamefont {Ranft}\ \emph {et~al.}(2010)\citenamefont {Ranft},
  \citenamefont {Basana}, \citenamefont {Elgeti}, \citenamefont {Joannya},
  \citenamefont {Prost},\ and\ \citenamefont {J{\"{u}}licher}}]{ranft2010}%
  \BibitemOpen
  \bibfield  {author} {\bibinfo {author} {\bibfnamefont {J.}~\bibnamefont
  {Ranft}}, \bibinfo {author} {\bibfnamefont {M.}~\bibnamefont {Basana}},
  \bibinfo {author} {\bibfnamefont {J.}~\bibnamefont {Elgeti}}, \bibinfo
  {author} {\bibfnamefont {J.~F.}\ \bibnamefont {Joannya}}, \bibinfo {author}
  {\bibfnamefont {J.}~\bibnamefont {Prost}}, \ and\ \bibinfo {author}
  {\bibfnamefont {F.}~\bibnamefont {J{\"{u}}licher}},\ }\href {\doibase
  10.1073/pnas.1011086107} {\bibfield  {journal} {\bibinfo  {journal} {Proc.
  Natl. Acad. Sci. USA}\ }\textbf {\bibinfo {volume} {107}},\ \bibinfo {pages}
  {20863} (\bibinfo {year} {2010})}\BibitemShut {NoStop}%
\bibitem [{\citenamefont {Cates}\ and\ \citenamefont
  {Tailleur}(2015)}]{cates2015}%
  \BibitemOpen
  \bibfield  {author} {\bibinfo {author} {\bibfnamefont {M.~E.}\ \bibnamefont
  {Cates}}\ and\ \bibinfo {author} {\bibfnamefont {J.}~\bibnamefont
  {Tailleur}},\ }\href {\doibase 10.1146/annurev-conmatphys-031214-014710}
  {\bibfield  {journal} {\bibinfo  {journal} {annu. Rev. Condens. Matter
  Phys.}\ }\textbf {\bibinfo {volume} {6}},\ \bibinfo {pages} {219} (\bibinfo
  {year} {2015})}\BibitemShut {NoStop}%
\bibitem [{\citenamefont {Prost}\ \emph {et~al.}(2015)\citenamefont {Prost},
  \citenamefont {J{\"{u}}licher},\ and\ \citenamefont {Joanny}}]{jacques15}%
  \BibitemOpen
  \bibfield  {author} {\bibinfo {author} {\bibfnamefont {J.}~\bibnamefont
  {Prost}}, \bibinfo {author} {\bibfnamefont {F.}~\bibnamefont
  {J{\"{u}}licher}}, \ and\ \bibinfo {author} {\bibfnamefont {J.-F.}\
  \bibnamefont {Joanny}},\ }\href {\doibase 10.1038/nphys3224} {\bibfield
  {journal} {\bibinfo  {journal} {Nat. Phys.}\ }\textbf {\bibinfo {volume}
  {11}},\ \bibinfo {pages} {111} (\bibinfo {year} {2015})}\BibitemShut
  {NoStop}%
\bibitem [{\citenamefont {Robin}\ \emph {et~al.}(2014)\citenamefont {Robin},
  \citenamefont {McFadden}, \citenamefont {Yao},\ and\ \citenamefont
  {Munro}}]{robin2014}%
  \BibitemOpen
  \bibfield  {author} {\bibinfo {author} {\bibfnamefont {F.~B.}\ \bibnamefont
  {Robin}}, \bibinfo {author} {\bibfnamefont {W.~M.}\ \bibnamefont {McFadden}},
  \bibinfo {author} {\bibfnamefont {B.}~\bibnamefont {Yao}}, \ and\ \bibinfo
  {author} {\bibfnamefont {E.~M.}\ \bibnamefont {Munro}},\ }\href {\doibase
  10.1038/nmeth.2928} {\bibfield  {journal} {\bibinfo  {journal} {Nat. Meth.}\
  }\textbf {\bibinfo {volume} {11}},\ \bibinfo {pages} {677} (\bibinfo {year}
  {2014})}\BibitemShut {NoStop}%
\bibitem [{\citenamefont {Fritzsche}\ \emph {et~al.}(2016)\citenamefont
  {Fritzsche}, \citenamefont {Erlenk{\"{a}}mper}, \citenamefont {Moeendarbary},
  \citenamefont {Charras},\ and\ \citenamefont {Kruse}}]{fritzsche2016}%
  \BibitemOpen
  \bibfield  {author} {\bibinfo {author} {\bibfnamefont {M.}~\bibnamefont
  {Fritzsche}}, \bibinfo {author} {\bibfnamefont {C.}~\bibnamefont
  {Erlenk{\"{a}}mper}}, \bibinfo {author} {\bibfnamefont {E.}~\bibnamefont
  {Moeendarbary}}, \bibinfo {author} {\bibfnamefont {G.}~\bibnamefont
  {Charras}}, \ and\ \bibinfo {author} {\bibfnamefont {K.}~\bibnamefont
  {Kruse}},\ }\href {\doibase 10.1126/sciadv.1501337} {\bibfield  {journal}
  {\bibinfo  {journal} {Sci. Adv.}\ }\textbf {\bibinfo {volume} {2}},\ \bibinfo
  {pages} {e1501337} (\bibinfo {year} {2016})}\BibitemShut {NoStop}%
\bibitem [{\citenamefont {Carlsson}(2010)}]{carlsson2010}%
  \BibitemOpen
  \bibfield  {author} {\bibinfo {author} {\bibfnamefont {A.~E.}\ \bibnamefont
  {Carlsson}},\ }\href {\doibase 10.1146/annurev.biophys.093008.131207}
  {\bibfield  {journal} {\bibinfo  {journal} {Annu. Rev. Biophys.}\ }\textbf
  {\bibinfo {volume} {6}},\ \bibinfo {pages} {91} (\bibinfo {year}
  {2010})}\BibitemShut {NoStop}%
\bibitem [{\citenamefont {McFadden}\ \emph {et~al.}(2017)\citenamefont
  {McFadden}, \citenamefont {McCall}, \citenamefont {Gardel},\ and\
  \citenamefont {Munro}}]{william2017}%
  \BibitemOpen
  \bibfield  {author} {\bibinfo {author} {\bibfnamefont {W.~M.}\ \bibnamefont
  {McFadden}}, \bibinfo {author} {\bibfnamefont {P.~M.}\ \bibnamefont
  {McCall}}, \bibinfo {author} {\bibfnamefont {M.~L.}\ \bibnamefont {Gardel}},
  \ and\ \bibinfo {author} {\bibfnamefont {E.~M.}\ \bibnamefont {Munro}},\
  }\href {\doibase 10.1371/journal.pcbi.1005811} {\bibfield  {journal}
  {\bibinfo  {journal} {PLOS: Comp. Biol.}\ }\textbf {\bibinfo {volume} {13}},\
  \bibinfo {pages} {e1005811} (\bibinfo {year} {2017})}\BibitemShut {NoStop}%
\bibitem [{\citenamefont {Das}(2004)}]{das2004}%
  \BibitemOpen
  \bibfield  {author} {\bibinfo {author} {\bibfnamefont {S.~P.}\ \bibnamefont
  {Das}},\ }\href {\doibase 10.1103/RevModPhys.76.785} {\bibfield  {journal}
  {\bibinfo  {journal} {Rev. Mod. Phys.}\ }\textbf {\bibinfo {volume} {76}},\
  \bibinfo {pages} {785} (\bibinfo {year} {2004})}\BibitemShut {NoStop}%
\bibitem [{\citenamefont {G{\"{o}}tze}(2008)}]{goetzebook}%
  \BibitemOpen
  \bibfield  {author} {\bibinfo {author} {\bibfnamefont {W.}~\bibnamefont
  {G{\"{o}}tze}},\ }\href@noop {} {\emph {\bibinfo {title} {Complex Dynamics of
  Glass-Forming Liquids: A Mode-Coupling Theory}}}\ (\bibinfo  {publisher}
  {Oxford University Press, Oxford},\ \bibinfo {year} {2008})\BibitemShut
  {NoStop}%
\bibitem [{\citenamefont {Fritzsche}\ \emph {et~al.}(2013)\citenamefont
  {Fritzsche}, \citenamefont {Lewalle}, \citenamefont {Duke}, \citenamefont
  {Kruse},\ and\ \citenamefont {Charras}}]{fritzsche2013}%
  \BibitemOpen
  \bibfield  {author} {\bibinfo {author} {\bibfnamefont {M.}~\bibnamefont
  {Fritzsche}}, \bibinfo {author} {\bibfnamefont {A.}~\bibnamefont {Lewalle}},
  \bibinfo {author} {\bibfnamefont {T.}~\bibnamefont {Duke}}, \bibinfo {author}
  {\bibfnamefont {K.}~\bibnamefont {Kruse}}, \ and\ \bibinfo {author}
  {\bibfnamefont {G.}~\bibnamefont {Charras}},\ }\href {\doibase
  10.1091/mbc.E12-06-0485} {\bibfield  {journal} {\bibinfo  {journal} {Mol.
  Biol. Cell}\ }\textbf {\bibinfo {volume} {24}},\ \bibinfo {pages} {757}
  (\bibinfo {year} {2013})}\BibitemShut {NoStop}%
\bibitem [{\citenamefont {Wollrab}\ \emph {et~al.}(2016)\citenamefont
  {Wollrab}, \citenamefont {Thiagarajan}, \citenamefont {Wald}, \citenamefont
  {Kruse},\ and\ \citenamefont {Riveline}}]{daniel2016}%
  \BibitemOpen
  \bibfield  {author} {\bibinfo {author} {\bibfnamefont {V.}~\bibnamefont
  {Wollrab}}, \bibinfo {author} {\bibfnamefont {R.}~\bibnamefont
  {Thiagarajan}}, \bibinfo {author} {\bibfnamefont {A.}~\bibnamefont {Wald}},
  \bibinfo {author} {\bibfnamefont {K.}~\bibnamefont {Kruse}}, \ and\ \bibinfo
  {author} {\bibfnamefont {D.}~\bibnamefont {Riveline}},\ }\href {\doibase
  10.1038/ncomms11860} {\bibfield  {journal} {\bibinfo  {journal} {Nat. Comm.}\
  }\textbf {\bibinfo {volume} {7}},\ \bibinfo {pages} {11860} (\bibinfo {year}
  {2016})}\BibitemShut {NoStop}%
\bibitem [{\citenamefont {Reymann}\ \emph {et~al.}(2012)\citenamefont
  {Reymann}, \citenamefont {Boujemaa-Paterski}, \citenamefont {Martiel},
  \citenamefont {Gu{\'{e}}rin}, \citenamefont {Cao}, \citenamefont {Chin},
  \citenamefont {Cruz}, \citenamefont {Th{\'{e}}ry},\ and\ \citenamefont
  {Blanchoin}}]{annececile2012}%
  \BibitemOpen
  \bibfield  {author} {\bibinfo {author} {\bibfnamefont {A.-C.}\ \bibnamefont
  {Reymann}}, \bibinfo {author} {\bibfnamefont {R.}~\bibnamefont
  {Boujemaa-Paterski}}, \bibinfo {author} {\bibfnamefont {J.-L.}\ \bibnamefont
  {Martiel}}, \bibinfo {author} {\bibfnamefont {C.}~\bibnamefont
  {Gu{\'{e}}rin}}, \bibinfo {author} {\bibfnamefont {W.}~\bibnamefont {Cao}},
  \bibinfo {author} {\bibfnamefont {H.~F.}\ \bibnamefont {Chin}}, \bibinfo
  {author} {\bibfnamefont {E.~M. D.~L.}\ \bibnamefont {Cruz}}, \bibinfo
  {author} {\bibfnamefont {M.}~\bibnamefont {Th{\'{e}}ry}}, \ and\ \bibinfo
  {author} {\bibfnamefont {L.}~\bibnamefont {Blanchoin}},\ }\href {\doibase
  10.1126/science.1221708} {\bibfield  {journal} {\bibinfo  {journal}
  {Science}\ }\textbf {\bibinfo {volume} {336}},\ \bibinfo {pages} {1310}
  (\bibinfo {year} {2012})}\BibitemShut {NoStop}%
\bibitem [{\citenamefont {Haviv}\ \emph {et~al.}(2008)\citenamefont {Haviv},
  \citenamefont {Gillo}, \citenamefont {Backouche},\ and\ \citenamefont
  {Bernheim-Groswasser}}]{haviv2008}%
  \BibitemOpen
  \bibfield  {author} {\bibinfo {author} {\bibfnamefont {L.}~\bibnamefont
  {Haviv}}, \bibinfo {author} {\bibfnamefont {D.}~\bibnamefont {Gillo}},
  \bibinfo {author} {\bibfnamefont {F.}~\bibnamefont {Backouche}}, \ and\
  \bibinfo {author} {\bibfnamefont {A.}~\bibnamefont {Bernheim-Groswasser}},\
  }\href {\doibase 10.1016/j.jmb.2007.09.066} {\bibfield  {journal} {\bibinfo
  {journal} {J. Mol. Biol.}\ }\textbf {\bibinfo {volume} {375}},\ \bibinfo
  {pages} {325} (\bibinfo {year} {2008})}\BibitemShut {NoStop}%
\bibitem [{\citenamefont {Wang}\ \emph {et~al.}(2011)\citenamefont {Wang},
  \citenamefont {Shen},\ and\ \citenamefont {Wolynes}}]{wangJCP2011}%
  \BibitemOpen
  \bibfield  {author} {\bibinfo {author} {\bibfnamefont {S.}~\bibnamefont
  {Wang}}, \bibinfo {author} {\bibfnamefont {T.}~\bibnamefont {Shen}}, \ and\
  \bibinfo {author} {\bibfnamefont {P.~G.}\ \bibnamefont {Wolynes}},\ }\href
  {\doibase 10.1063/1.3518450} {\bibfield  {journal} {\bibinfo  {journal} {J.
  Chem. Phys.}\ }\textbf {\bibinfo {volume} {134}},\ \bibinfo {pages} {014510}
  (\bibinfo {year} {2011})}\BibitemShut {NoStop}%
\bibitem [{\citenamefont {Nandi}\ and\ \citenamefont {Gov}(2017)}]{noneqmct}%
  \BibitemOpen
  \bibfield  {author} {\bibinfo {author} {\bibfnamefont {S.~K.}\ \bibnamefont
  {Nandi}}\ and\ \bibinfo {author} {\bibfnamefont {N.~S.}\ \bibnamefont
  {Gov}},\ }\href {\doibase 10.1039/C7SM01648D} {\bibfield  {journal} {\bibinfo
   {journal} {Soft Matter}\ }\textbf {\bibinfo {volume} {13}},\ \bibinfo
  {pages} {7609} (\bibinfo {year} {2017})}\BibitemShut {NoStop}%
\bibitem [{\citenamefont {Shen}\ and\ \citenamefont
  {Wolynes}(2004)}]{shen2004}%
  \BibitemOpen
  \bibfield  {author} {\bibinfo {author} {\bibfnamefont {T.}~\bibnamefont
  {Shen}}\ and\ \bibinfo {author} {\bibfnamefont {P.~G.}\ \bibnamefont
  {Wolynes}},\ }\href {\doibase 10.1073/pnas.0402602101} {\bibfield  {journal}
  {\bibinfo  {journal} {Proc. Natl. Acad. Sci. (USA)}\ }\textbf {\bibinfo
  {volume} {101}},\ \bibinfo {pages} {8547} (\bibinfo {year}
  {2004})}\BibitemShut {NoStop}%
\bibitem [{\citenamefont {Lu}\ \emph {et~al.}(2006)\citenamefont {Lu},
  \citenamefont {Hasty},\ and\ \citenamefont {Wolynes}}]{lu2006}%
  \BibitemOpen
  \bibfield  {author} {\bibinfo {author} {\bibfnamefont {T.}~\bibnamefont
  {Lu}}, \bibinfo {author} {\bibfnamefont {J.}~\bibnamefont {Hasty}}, \ and\
  \bibinfo {author} {\bibfnamefont {P.~G.}\ \bibnamefont {Wolynes}},\ }\href
  {\doibase 10.1529/biophysj.105.074914} {\bibfield  {journal} {\bibinfo
  {journal} {Biophys. J.}\ }\textbf {\bibinfo {volume} {91}},\ \bibinfo {pages}
  {84} (\bibinfo {year} {2006})}\BibitemShut {NoStop}%
\bibitem [{\citenamefont {Wang}\ and\ \citenamefont
  {Wolynes}(2013)}]{wang2013}%
  \BibitemOpen
  \bibfield  {author} {\bibinfo {author} {\bibfnamefont {S.}~\bibnamefont
  {Wang}}\ and\ \bibinfo {author} {\bibfnamefont {P.~G.}\ \bibnamefont
  {Wolynes}},\ }\href {\doibase 10.1063/1.4773349} {\bibfield  {journal}
  {\bibinfo  {journal} {J. Chem. Phys.}\ }\textbf {\bibinfo {volume} {138}},\
  \bibinfo {pages} {12A521} (\bibinfo {year} {2013})}\BibitemShut {NoStop}%
\bibitem [{\citenamefont {Loi}\ \emph {et~al.}(2011{\natexlab{a}})\citenamefont
  {Loi}, \citenamefont {Mossa},\ and\ \citenamefont {Cugliandolo}}]{loi2011}%
  \BibitemOpen
  \bibfield  {author} {\bibinfo {author} {\bibfnamefont {D.}~\bibnamefont
  {Loi}}, \bibinfo {author} {\bibfnamefont {S.}~\bibnamefont {Mossa}}, \ and\
  \bibinfo {author} {\bibfnamefont {L.~F.}\ \bibnamefont {Cugliandolo}},\
  }\href {\doibase 10.1039/C0SM01484B} {\bibfield  {journal} {\bibinfo
  {journal} {Soft Matter}\ }\textbf {\bibinfo {volume} {7}},\ \bibinfo {pages}
  {3726} (\bibinfo {year} {2011}{\natexlab{a}})}\BibitemShut {NoStop}%
\bibitem [{\citenamefont {Loi}\ \emph {et~al.}(2011{\natexlab{b}})\citenamefont
  {Loi}, \citenamefont {Mossa},\ and\ \citenamefont {Cugliandolo}}]{loi2011a}%
  \BibitemOpen
  \bibfield  {author} {\bibinfo {author} {\bibfnamefont {D.}~\bibnamefont
  {Loi}}, \bibinfo {author} {\bibfnamefont {S.}~\bibnamefont {Mossa}}, \ and\
  \bibinfo {author} {\bibfnamefont {L.~F.}\ \bibnamefont {Cugliandolo}},\
  }\href {\doibase 10.1039/C1SM05819C} {\bibfield  {journal} {\bibinfo
  {journal} {Soft Matter}\ }\textbf {\bibinfo {volume} {7}},\ \bibinfo {pages}
  {10193} (\bibinfo {year} {2011}{\natexlab{b}})}\BibitemShut {NoStop}%
\bibitem [{\citenamefont {Cugliandolo}(2011)}]{cugliandolo2011}%
  \BibitemOpen
  \bibfield  {author} {\bibinfo {author} {\bibfnamefont {L.~F.}\ \bibnamefont
  {Cugliandolo}},\ }\href {\doibase 10.1088/1751-8113/44/48/483001} {\bibfield
  {journal} {\bibinfo  {journal} {J. Phys. A: Math. Theor.}\ }\textbf {\bibinfo
  {volume} {44}},\ \bibinfo {pages} {483001} (\bibinfo {year}
  {2011})}\BibitemShut {NoStop}%
\bibitem [{\citenamefont {Suma}\ \emph {et~al.}(2014)\citenamefont {Suma},
  \citenamefont {Gonnella}, \citenamefont {Laghezza}, \citenamefont {Lamura},
  \citenamefont {Mossa},\ and\ \citenamefont {Cugliandolo}}]{suma2014}%
  \BibitemOpen
  \bibfield  {author} {\bibinfo {author} {\bibfnamefont {A.}~\bibnamefont
  {Suma}}, \bibinfo {author} {\bibfnamefont {G.}~\bibnamefont {Gonnella}},
  \bibinfo {author} {\bibfnamefont {G.}~\bibnamefont {Laghezza}}, \bibinfo
  {author} {\bibfnamefont {A.}~\bibnamefont {Lamura}}, \bibinfo {author}
  {\bibfnamefont {A.}~\bibnamefont {Mossa}}, \ and\ \bibinfo {author}
  {\bibfnamefont {L.~F.}\ \bibnamefont {Cugliandolo}},\ }\href {\doibase
  10.1103/PhysRevE.90.052130} {\bibfield  {journal} {\bibinfo  {journal} {Phys.
  Rev. E}\ }\textbf {\bibinfo {volume} {90}},\ \bibinfo {pages} {052130}
  (\bibinfo {year} {2014})}\BibitemShut {NoStop}%
\bibitem [{\citenamefont {Ben-Isaac}\ \emph {et~al.}(2015)\citenamefont
  {Ben-Isaac}, \citenamefont {Fodor}, \citenamefont {Visco}, \citenamefont {van
  Wijland},\ and\ \citenamefont {Gov}}]{benisaac2015}%
  \BibitemOpen
  \bibfield  {author} {\bibinfo {author} {\bibfnamefont {E.}~\bibnamefont
  {Ben-Isaac}}, \bibinfo {author} {\bibfnamefont {{\'{E}}.}~\bibnamefont
  {Fodor}}, \bibinfo {author} {\bibfnamefont {P.}~\bibnamefont {Visco}},
  \bibinfo {author} {\bibfnamefont {F.}~\bibnamefont {van Wijland}}, \ and\
  \bibinfo {author} {\bibfnamefont {N.~S.}\ \bibnamefont {Gov}},\ }\href
  {\doibase 10.1103/PhysRevE.92.012716} {\bibfield  {journal} {\bibinfo
  {journal} {Phys. Rev. E}\ }\textbf {\bibinfo {volume} {92}},\ \bibinfo
  {pages} {012716} (\bibinfo {year} {2015})}\BibitemShut {NoStop}%
\bibitem [{\citenamefont {Zaccarelli}\ \emph {et~al.}(2002)\citenamefont
  {Zaccarelli}, \citenamefont {Foffi}, \citenamefont {Gregorio}, \citenamefont
  {Sciortino}, \citenamefont {Tartaglia},\ and\ \citenamefont
  {Dawson1}}]{zaccarelli2002}%
  \BibitemOpen
  \bibfield  {author} {\bibinfo {author} {\bibfnamefont {E.}~\bibnamefont
  {Zaccarelli}}, \bibinfo {author} {\bibfnamefont {G.}~\bibnamefont {Foffi}},
  \bibinfo {author} {\bibfnamefont {P.~D.}\ \bibnamefont {Gregorio}}, \bibinfo
  {author} {\bibfnamefont {F.}~\bibnamefont {Sciortino}}, \bibinfo {author}
  {\bibfnamefont {P.}~\bibnamefont {Tartaglia}}, \ and\ \bibinfo {author}
  {\bibfnamefont {K.~A.}\ \bibnamefont {Dawson1}},\ }\href
  {http://stacks.iop.org/0953-8984/14/i=9/a=330} {\bibfield  {journal}
  {\bibinfo  {journal} {J. Phys.: Condens. Matter}\ }\textbf {\bibinfo {volume}
  {14}},\ \bibinfo {pages} {2413} (\bibinfo {year} {2002})}\BibitemShut
  {NoStop}%
\bibitem [{\citenamefont {Zwanzig}(2001)}]{zwanzigbook}%
  \BibitemOpen
  \bibfield  {author} {\bibinfo {author} {\bibfnamefont {R.}~\bibnamefont
  {Zwanzig}},\ }\href@noop {} {\emph {\bibinfo {title} {Nonequilibrium
  Statistical Mechanics}}}\ (\bibinfo  {publisher} {Oxford University Press},\
  \bibinfo {year} {2001})\BibitemShut {NoStop}%
\bibitem [{\citenamefont {Ramakrishnan}\ and\ \citenamefont
  {Yussouff}(1979)}]{ramakrishnan1979}%
  \BibitemOpen
  \bibfield  {author} {\bibinfo {author} {\bibfnamefont {T.~V.}\ \bibnamefont
  {Ramakrishnan}}\ and\ \bibinfo {author} {\bibfnamefont {M.}~\bibnamefont
  {Yussouff}},\ }\href {\doibase 10.1103/PhysRevB.19.2775} {\bibfield
  {journal} {\bibinfo  {journal} {Phys. Rev. B}\ }\textbf {\bibinfo {volume}
  {19}},\ \bibinfo {pages} {2775} (\bibinfo {year} {1979})}\BibitemShut
  {NoStop}%
\bibitem [{\citenamefont {Nandi}(2015)}]{saroj2015}%
  \BibitemOpen
  \bibfield  {author} {\bibinfo {author} {\bibfnamefont {S.~K.}\ \bibnamefont
  {Nandi}},\ }\href {\doibase 10.1103/PhysRevE.92.042306} {\bibfield  {journal}
  {\bibinfo  {journal} {Phys. Rev. E}\ }\textbf {\bibinfo {volume} {92}},\
  \bibinfo {pages} {042306} (\bibinfo {year} {2015})}\BibitemShut {NoStop}%
\bibitem [{Note1()}]{Note1}%
  \BibitemOpen
  \bibinfo {note} {Note that even in a passive system it is not clear what
  should be the correct form of $c({\protect \bf {r}})$ (see
  Appendix).}\BibitemShut {Stop}%
\bibitem [{\citenamefont {Jungbauer}\ \emph {et~al.}(2008)\citenamefont
  {Jungbauer}, \citenamefont {Gao}, \citenamefont {Spatz},\ and\ \citenamefont
  {Kemkemer}}]{jungbauer2008}%
  \BibitemOpen
  \bibfield  {author} {\bibinfo {author} {\bibfnamefont {S.}~\bibnamefont
  {Jungbauer}}, \bibinfo {author} {\bibfnamefont {H.}~\bibnamefont {Gao}},
  \bibinfo {author} {\bibfnamefont {J.~P.}\ \bibnamefont {Spatz}}, \ and\
  \bibinfo {author} {\bibfnamefont {R.}~\bibnamefont {Kemkemer}},\ }\href
  {\doibase 10.1529/biophysj.107.128611} {\bibfield  {journal} {\bibinfo
  {journal} {Biophys. J.}\ }\textbf {\bibinfo {volume} {95}},\ \bibinfo {pages}
  {3470} (\bibinfo {year} {2008})}\BibitemShut {NoStop}%
\bibitem [{\citenamefont {Hansen}\ and\ \citenamefont
  {McDonald}(2013)}]{hansenmcdonald}%
  \BibitemOpen
  \bibfield  {author} {\bibinfo {author} {\bibfnamefont {J.~P.}\ \bibnamefont
  {Hansen}}\ and\ \bibinfo {author} {\bibfnamefont {I.~R.}\ \bibnamefont
  {McDonald}},\ }\href@noop {} {\emph {\bibinfo {title} {Theory of Simple
  Liquids}}},\ \bibinfo {edition} {4th}\ ed.\ (\bibinfo  {publisher}
  {Elsevier},\ \bibinfo {year} {2013})\BibitemShut {NoStop}%
\bibitem [{\citenamefont {Leutheusser}(1984)}]{leutheusser1984}%
  \BibitemOpen
  \bibfield  {author} {\bibinfo {author} {\bibfnamefont {E.}~\bibnamefont
  {Leutheusser}},\ }\href {\doibase 10.1103/PhysRevA.29.2765} {\bibfield
  {journal} {\bibinfo  {journal} {Phys. Rev. A}\ }\textbf {\bibinfo {volume}
  {29}},\ \bibinfo {pages} {2765} (\bibinfo {year} {1984})}\BibitemShut
  {NoStop}%
\bibitem [{\citenamefont {Kirkpatrick}(1985)}]{kirkpatrick1985}%
  \BibitemOpen
  \bibfield  {author} {\bibinfo {author} {\bibfnamefont {T.~R.}\ \bibnamefont
  {Kirkpatrick}},\ }\href {\doibase 10.1103/PhysRevA.31.939} {\bibfield
  {journal} {\bibinfo  {journal} {Phys. Rev. A}\ }\textbf {\bibinfo {volume}
  {31}},\ \bibinfo {pages} {939} (\bibinfo {year} {1985})}\BibitemShut
  {NoStop}%
\bibitem [{\citenamefont {Brader}\ \emph {et~al.}(2009)\citenamefont {Brader},
  \citenamefont {Voigtmann}, \citenamefont {Fuchs}, \citenamefont {Larson},\
  and\ \citenamefont {Cates}}]{brader2009}%
  \BibitemOpen
  \bibfield  {author} {\bibinfo {author} {\bibfnamefont {J.~M.}\ \bibnamefont
  {Brader}}, \bibinfo {author} {\bibfnamefont {T.}~\bibnamefont {Voigtmann}},
  \bibinfo {author} {\bibfnamefont {M.}~\bibnamefont {Fuchs}}, \bibinfo
  {author} {\bibfnamefont {R.~G.}\ \bibnamefont {Larson}}, \ and\ \bibinfo
  {author} {\bibfnamefont {M.~E.}\ \bibnamefont {Cates}},\ }\href {\doibase
  10.1073/pnas.0905330106} {\bibfield  {journal} {\bibinfo  {journal} {Proc.
  Natl. Acad. Sci. (USA)}\ }\textbf {\bibinfo {volume} {106}},\ \bibinfo
  {pages} {15186} (\bibinfo {year} {2009})}\BibitemShut {NoStop}%
\bibitem [{Note2()}]{Note2}%
  \BibitemOpen
  \bibinfo {note} {Within MCT, the relaxation dynamics at all wave vectors are
  similar \cite {das2004}, therefore, writing the theory for any other wave
  vector is equivalent.}\BibitemShut {Stop}%
\bibitem [{\citenamefont {Singh}\ \emph {et~al.}(1998)\citenamefont {Singh},
  \citenamefont {Li}, \citenamefont {G{\"{o}}tze}, \citenamefont {Fuchs},\ and\
  \citenamefont {Cummins}}]{sing1998}%
  \BibitemOpen
  \bibfield  {author} {\bibinfo {author} {\bibfnamefont {A.~P.}\ \bibnamefont
  {Singh}}, \bibinfo {author} {\bibfnamefont {G.}~\bibnamefont {Li}}, \bibinfo
  {author} {\bibfnamefont {W.}~\bibnamefont {G{\"{o}}tze}}, \bibinfo {author}
  {\bibfnamefont {M.}~\bibnamefont {Fuchs}}, \ and\ \bibinfo {author}
  {\bibfnamefont {T.~F. H.~Z.}\ \bibnamefont {Cummins}},\ }\href {\doibase
  10.1016/S0022-3093(98)00583-3} {\bibfield  {journal} {\bibinfo  {journal} {J.
  Non-crystalline Solids}\ }\textbf {\bibinfo {volume} {235-237}},\ \bibinfo
  {pages} {66} (\bibinfo {year} {1998})}\BibitemShut {NoStop}%
\bibitem [{\citenamefont {G{\"{o}}tze}\ and\ \citenamefont
  {Mayr}(2000)}]{goetze2000}%
  \BibitemOpen
  \bibfield  {author} {\bibinfo {author} {\bibfnamefont {W.}~\bibnamefont
  {G{\"{o}}tze}}\ and\ \bibinfo {author} {\bibfnamefont {M.~R.}\ \bibnamefont
  {Mayr}},\ }\href {\doibase 10.1103/PhysRevE.61.587} {\bibfield  {journal}
  {\bibinfo  {journal} {Phys. Rev. E}\ }\textbf {\bibinfo {volume} {61}},\
  \bibinfo {pages} {587} (\bibinfo {year} {2000})}\BibitemShut {NoStop}%
\bibitem [{\citenamefont {Berthier}\ and\ \citenamefont
  {Kurchan}(2013)}]{berthier2013}%
  \BibitemOpen
  \bibfield  {author} {\bibinfo {author} {\bibfnamefont {L.}~\bibnamefont
  {Berthier}}\ and\ \bibinfo {author} {\bibfnamefont {J.}~\bibnamefont
  {Kurchan}},\ }\href {\doibase 10.1038/nphys2592} {\bibfield  {journal}
  {\bibinfo  {journal} {Nat. Phys.}\ }\textbf {\bibinfo {volume} {9}},\
  \bibinfo {pages} {310} (\bibinfo {year} {2013})}\BibitemShut {NoStop}%
\bibitem [{\citenamefont {Mandal}\ \emph {et~al.}(2016)\citenamefont {Mandal},
  \citenamefont {Bhuyan}, \citenamefont {Rao},\ and\ \citenamefont
  {Dasgupta}}]{mandal2016}%
  \BibitemOpen
  \bibfield  {author} {\bibinfo {author} {\bibfnamefont {R.}~\bibnamefont
  {Mandal}}, \bibinfo {author} {\bibfnamefont {P.~J.}\ \bibnamefont {Bhuyan}},
  \bibinfo {author} {\bibfnamefont {M.}~\bibnamefont {Rao}}, \ and\ \bibinfo
  {author} {\bibfnamefont {C.}~\bibnamefont {Dasgupta}},\ }\href {\doibase
  10.1039/c5sm02950c} {\bibfield  {journal} {\bibinfo  {journal} {Soft Matter}\
  }\textbf {\bibinfo {volume} {12}},\ \bibinfo {pages} {6268} (\bibinfo {year}
  {2016})}\BibitemShut {NoStop}%
\bibitem [{\citenamefont {Flenner}\ \emph {et~al.}(2016)\citenamefont
  {Flenner}, \citenamefont {Szamel},\ and\ \citenamefont
  {Berthier}}]{flenner2016}%
  \BibitemOpen
  \bibfield  {author} {\bibinfo {author} {\bibfnamefont {E.}~\bibnamefont
  {Flenner}}, \bibinfo {author} {\bibfnamefont {G.}~\bibnamefont {Szamel}}, \
  and\ \bibinfo {author} {\bibfnamefont {L.}~\bibnamefont {Berthier}},\ }\href
  {\doibase 10.1039/c6sm01322h} {\bibfield  {journal} {\bibinfo  {journal}
  {Soft Matter}\ }\textbf {\bibinfo {volume} {12}},\ \bibinfo {pages} {7136}
  (\bibinfo {year} {2016})}\BibitemShut {NoStop}%
\bibitem [{\citenamefont {Nandi}(2018)}]{viscmct}%
  \BibitemOpen
  \bibfield  {author} {\bibinfo {author} {\bibfnamefont {S.~K.}\ \bibnamefont
  {Nandi}},\ }\href {https://arxiv.org/abs/1607.04478} {\bibfield  {journal}
  {\bibinfo  {journal} {arXiv: 1607.04478}\ } (\bibinfo {year}
  {2018})}\BibitemShut {NoStop}%
\bibitem [{\citenamefont {Nandi}\ \emph {et~al.}(2016)\citenamefont {Nandi},
  \citenamefont {Biroli},\ and\ \citenamefont {Tarjus}}]{rfimmct}%
  \BibitemOpen
  \bibfield  {author} {\bibinfo {author} {\bibfnamefont {S.~K.}\ \bibnamefont
  {Nandi}}, \bibinfo {author} {\bibfnamefont {G.}~\bibnamefont {Biroli}}, \
  and\ \bibinfo {author} {\bibfnamefont {G.}~\bibnamefont {Tarjus}},\ }\href
  {\doibase 10.1103/PhysRevLett.116.145701} {\bibfield  {journal} {\bibinfo
  {journal} {Phys. Rev. Lett.}\ }\textbf {\bibinfo {volume} {116}},\ \bibinfo
  {pages} {145701} (\bibinfo {year} {2016})}\BibitemShut {NoStop}%
\bibitem [{\citenamefont {Kubo}(1966)}]{kubo1966}%
  \BibitemOpen
  \bibfield  {author} {\bibinfo {author} {\bibfnamefont {R.}~\bibnamefont
  {Kubo}},\ }\href {\doibase 10.1088/0034-4885/29/1/306} {\bibfield  {journal}
  {\bibinfo  {journal} {Rep. Prog. Phys.}\ }\textbf {\bibinfo {volume} {29}},\
  \bibinfo {pages} {255} (\bibinfo {year} {1966})}\BibitemShut {NoStop}%
\bibitem [{\citenamefont {Amit}\ and\ \citenamefont {Mayor}(2005)}]{amitbook}%
  \BibitemOpen
  \bibfield  {author} {\bibinfo {author} {\bibfnamefont {D.~J.}\ \bibnamefont
  {Amit}}\ and\ \bibinfo {author} {\bibfnamefont {V.~M.}\ \bibnamefont
  {Mayor}},\ }\href@noop {} {\emph {\bibinfo {title} {Field Theory, The
  Renormalization Group, And Critical Phenomena: Graphs To Computers}}}\
  (\bibinfo  {publisher} {World Scientific Publishing Co. Pvt. Ltd.,
  Singapore},\ \bibinfo {year} {2005})\BibitemShut {NoStop}%
\bibitem [{\citenamefont {Berthier}\ and\ \citenamefont
  {Biroli}(2011)}]{giulioreview}%
  \BibitemOpen
  \bibfield  {author} {\bibinfo {author} {\bibfnamefont {L.}~\bibnamefont
  {Berthier}}\ and\ \bibinfo {author} {\bibfnamefont {G.}~\bibnamefont
  {Biroli}},\ }\href {\doibase 10.1103/RevModPhys.83.587} {\bibfield  {journal}
  {\bibinfo  {journal} {Rev. Mod. Phys.}\ }\textbf {\bibinfo {volume} {83}},\
  \bibinfo {pages} {587} (\bibinfo {year} {2011})}\BibitemShut {NoStop}%
\bibitem [{\citenamefont {Brangwynne}\ \emph {et~al.}(2011)\citenamefont
  {Brangwynne}, \citenamefont {Mitchison},\ and\ \citenamefont
  {Hyman}}]{brangwynne2011}%
  \BibitemOpen
  \bibfield  {author} {\bibinfo {author} {\bibfnamefont {C.~P.}\ \bibnamefont
  {Brangwynne}}, \bibinfo {author} {\bibfnamefont {T.~J.}\ \bibnamefont
  {Mitchison}}, \ and\ \bibinfo {author} {\bibfnamefont {A.~A.}\ \bibnamefont
  {Hyman}},\ }\href {\doibase 10.1073/pnas.1017150108} {\bibfield  {journal}
  {\bibinfo  {journal} {Proc. Natl. Acad. Sci. (USA)}\ }\textbf {\bibinfo
  {volume} {108}},\ \bibinfo {pages} {4334} (\bibinfo {year}
  {2011})}\BibitemShut {NoStop}%
\bibitem [{\citenamefont {F{\"{u}}rthauer}\ \emph {et~al.}(2013)\citenamefont
  {F{\"{u}}rthauer}, \citenamefont {Strempel}, \citenamefont {Grill},\ and\
  \citenamefont {J{\"{u}}licher}}]{sebastian2013}%
  \BibitemOpen
  \bibfield  {author} {\bibinfo {author} {\bibfnamefont {S.}~\bibnamefont
  {F{\"{u}}rthauer}}, \bibinfo {author} {\bibfnamefont {M.}~\bibnamefont
  {Strempel}}, \bibinfo {author} {\bibfnamefont {S.~W.}\ \bibnamefont {Grill}},
  \ and\ \bibinfo {author} {\bibfnamefont {F.}~\bibnamefont {J{\"{u}}licher}},\
  }\href {\doibase 10.1103/PhysRevLett.110.048103} {\bibfield  {journal}
  {\bibinfo  {journal} {Phys. Rev. Lett.}\ }\textbf {\bibinfo {volume} {110}},\
  \bibinfo {pages} {048103} (\bibinfo {year} {2013})}\BibitemShut {NoStop}%
\bibitem [{\citenamefont {Szamel}(2016)}]{szamel2016}%
  \BibitemOpen
  \bibfield  {author} {\bibinfo {author} {\bibfnamefont {G.}~\bibnamefont
  {Szamel}},\ }\href {\doibase 10.1103/PhysRevE.93.012603} {\bibfield
  {journal} {\bibinfo  {journal} {Phys. Rev. E}\ }\textbf {\bibinfo {volume}
  {93}},\ \bibinfo {pages} {012603} (\bibinfo {year} {2016})}\BibitemShut
  {NoStop}%
\bibitem [{\citenamefont {Kawasaki}(2002)}]{kawasaki2002}%
  \BibitemOpen
  \bibfield  {author} {\bibinfo {author} {\bibfnamefont {K.}~\bibnamefont
  {Kawasaki}},\ }\href {\doibase 10.1023/A:1022161330306} {\bibfield  {journal}
  {\bibinfo  {journal} {J. Stat. Phys.}\ }\textbf {\bibinfo {volume} {110}},\
  \bibinfo {pages} {1249} (\bibinfo {year} {2002})}\BibitemShut {NoStop}%
\bibitem [{\citenamefont {Nandi}\ \emph {et~al.}(2011)\citenamefont {Nandi},
  \citenamefont {Bhattacharyya},\ and\ \citenamefont {Ramaswamy}}]{saroj2011}%
  \BibitemOpen
  \bibfield  {author} {\bibinfo {author} {\bibfnamefont {S.~K.}\ \bibnamefont
  {Nandi}}, \bibinfo {author} {\bibfnamefont {S.~M.}\ \bibnamefont
  {Bhattacharyya}}, \ and\ \bibinfo {author} {\bibfnamefont {S.}~\bibnamefont
  {Ramaswamy}},\ }\href {\doibase 10.1103/PhysRevE.84.061501} {\bibfield
  {journal} {\bibinfo  {journal} {Phys. Rev. E}\ }\textbf {\bibinfo {volume}
  {84}},\ \bibinfo {pages} {061501} (\bibinfo {year} {2011})}\BibitemShut
  {NoStop}%
\end{thebibliography}


%

\end{document}